\documentclass[twocolumn,epjc3]{svjour3}
\usepackage{amsmath}
\usepackage{amssymb}
\usepackage{bbm}
\RequirePackage[T1]{fontenc}
\smartqed
\RequirePackage{graphicx}
\usepackage{epstopdf}
\RequirePackage{mathptmx}
\RequirePackage{flushend}
\RequirePackage[numbers,sort&compress]{natbib}
\RequirePackage[colorlinks,citecolor=blue,urlcolor=blue,linkcolor=blue]{hyperref}

\journalname{Eur. Phys. J. C}

\begin{document}

\title{
  On the 2D Dirac oscillator in the presence of vector and scalar
  potentials in the cosmic string spacetime in the context of spin and
  pseudospin symmetries
}

\author{
  Daniel F. Lima\thanksref{e1,addr1}
  \and
  Fabiano M. Andrade\thanksref{e2,addr2}
  \and
  Luis B. Castro\thanksref{e3,addr1}
  \and
  Cleverson Filgueiras\thanksref{e4,addr3}
  \and
  Edilberto O. Silva\thanksref{e5,addr1}
}

\thankstext{e1}{e-mail: daniel.franca.lima@gmail.com}
\thankstext{e2}{e-mail: fmandrade@uepg.br}
\thankstext{e3}{e-mail: lrb.castro@ufma.br}
\thankstext{e4}{e-mail: cleversonfilgueiras@yahoo.com.br}
\thankstext{e5}{e-mail: edilbertoo@gmail.com}

\institute{%
  Departamento de F\'{i}sica,
  Universidade Federal do Maranh\~{a}o,
  65085-580 S\~{a}o Lu\'{i}s, MA, Brazil\label{addr1}
  \and
  Departamento de Matem\'{a}tica e Estat\'{i}stica,
  Universidade Estadual de Ponta Grossa,
  84030-900 Ponta Grossa, PR, Brazil\label{addr2}
  \and
  Departamento de F\'{i}sica,
  Universidade Federal de Lavras,
  Caixa Postal 3037,
  37200-000, Lavras, MG, Brazil\label{addr3}
}

\date{Received: 01 December 2018 / Accepted: 05 July 2019}

\maketitle

\begin{abstract}
The Dirac equation with both scalar and vector couplings describing the
dynamics of a two-dimensional Dirac oscillator in the cosmic string
spacetime is considered.
We derive the Dirac-Pauli equation and solve it in the limit of the spin
and the pseudo-spin symmetries.
We analyze the presence of cylindrical symmetric scalar potentials which
allows us to provide analytic solutions for the resultant field
equation.
By using an appropriate ansatz, we find that the radial equation is a
biconfluent Heun-like differential equation.
The solution of this equation provides us with more than one expression for
the energy eigenvalues of the oscillator.
We investigate these energies and find that there is a quantum condition
between them.
We study this condition in detail and find that it requires the fixation
of one of the physical parameters involved in the problem.
Expressions for the energy of the oscillator are obtained for some
values of the quantum number $n$.
Some particular cases which lead to known physical systems are also
addressed.
\end{abstract}

\section{Introduction}
\label{sec:introduction}

The study of the relativistic quantum dynamics of particles including
electromagnetic interactions is an usual framework for studying properties
of various physical systems.
The mechanism used to describe these systems is a natural generalization
of the coupling used in classical nonrelativistic quantum theory
\cite{Book.2012.Itzykson}.
This coupling is implemented, for charged particles with charge $e$,
through the so-called minimal coupling prescription, given in terms of
the modification of the $4$-momentum operator,
$p_{\mu }\rightarrow p_{\mu }-eA_{\mu }=\left( p_{0}-eA_{0},
\mathbf{p}-e\mathbf{A}\right)$, where $A_{\mu }=\left( V(r) ,-
\mathbf{A}\right) $ (with $\mathbf{A}$ being the vector potential and
$V(r)$ being the scalar potential) represents the $4$-vector potential
of the associated electromagnetic field.
This transformation preserves the gauge invariance associated with the
Maxwell's equations.
Another way to insert interaction in the dynamics of the particle is by
including a scalar potential through a modification in the mass term as
$M\rightarrow M+S(r)$.
In this realization, the potential $S \left(r\right)$ is coupled  like a
scalar, different from the minimum prescription, where the potential is
coupled as a time-like component of a 4-vector.
Although there is some similarity between the scalar and vector
couplings, they have different physical implications.
Actually, the scalar coupling acts equally on particles and
antiparticles. On the other hand, the vector coupling acts differently
on particles and antiparticles.
As a result, the energy of particle and antiparticle are not equals, so
that bound states exist only for one of the two kinds of particles
\cite{Book.2000.Greiner}.

Interesting issues that should be investigated with the insertion of the
couplings in the Dirac equation are the so-called the spin and the pseudo-spin
symmetries \cite{PR.2005.414.165}.
Basically, these symmetries occur when
the couplings are composed by a vector $V(r) $ and a scalar $
S(r) $ potential, under the assumption that $S(r)
=V(r) $ ($S(r) =-V(r) $), which is the
necessary condition for occurrence of exact spin (pseudo-spin) symmetry.
The
spin symmetry has been identified by studying heavy-light mesons \cite
{PRL.2001.86.204}, single antinucleon spectra \cite{PRL.2003.91.262501} and
dynamics of a light quark (antiquark) in the field of a heavy antiquark
(quark) \cite{PR.2005.414.165} while that the pseudo-spin symmetry occurs in the
motion of nucleons \cite{PR.1999.315.231,PR.2005.414.165}.
In recent studies, both the spin and the pseudo-spin symmetries appear in several aspects concerning, for instance, the supersymmetry \cite
{PLB.2011.699.309,JPA.2006.39.7737}, the Hartree-Fock theory \cite
{PLB.2006.639.242}, the electrons in graphene \cite{PRA.2015.92.062137} and
the interaction with a class of scalar and vector potentials \cite
{AoP.2017.378.88,AoP.2016.364.99,EPJC.2015.75.321,AoP.2015.356.83,CTP.2015.64.637,JTAP.2015.9.15,IJMPA.2016.311650190,EPL.2007.77.20009}.

An important physical system that can be studied by including such terms
of interactions in the Dirac equation is the Dirac oscillator
\cite{JPA.1989.22.817} (for a detailed description of
this model see Ref. \cite{Book.1998.Strange}).
The Dirac oscillator is a kind of tensor coupling with a linear
potential which in the nonrelativistic limit leads to the simple
harmonic oscillator with a strong spin-orbit coupling.
It was realized experimentally for the first time in 2013 by
Franco-Villafa\~{n}e \textit{et al.} in \cite{PRL.2013.111.170405}.
The Dirac oscillator is considered a natural model for studying
properties of physical systems because it is exactly soluble. In the
last years, several research have been developed in the context of this
theoretical framework.
For instance, it appears in the literature in the context of
mathematical physics 
\cite{AoP.2014.351.13,JPA.2006.39.5125,EPL.2014.108.30003,
PLA.2004.325.21,JPA.1997.30.2585,JPA.2005.38.1747,
JPA.1991.24.667,PLB.2012.710.478}, nuclear physics
\cite{PRC.2006.73.054309,PLA.2012.376.3475,PRC.2012.85.054617,
AP.2005.320.71}, quantum optics
\cite{JOB.2002.4.R1,OL.2010.35.1302,EPJB.2012.85.237,PRA.2007.76.041801},
supersymmetry \cite{JPA.1995.28.6447,JPA.2006.39.10909,CTP.2008.49.319},
theory of quantum deformations \cite{PLB.2014.731.327,PLB.2014.738.44}
and noncommutativity
\cite{PLA.2012.376.2467,IJTP.2013.52.441,IJTP.2012.51.2143,
IJMPA.2011.26.4991}.
Moreover, the Dirac oscillator embedded in a cosmic string background
has inspired a great deal of research in last years
\cite{EPJP.2018.133.409,EPJP.2012.127.82,GRG.2013.45.1847,PLA.2012.376.1269,
PRA.2011.84.32109,AoP.2013.336.489,NPB.1989.328.140,PRL.1989.62.1071,
PLA.2007.361.13,EPJC.2014.74.3187}.

In this work, we analyze in details the solutions of the Dirac
equation with both scalar and vector interactions under the
spin and the pseudo-spin symmetry limits in the cosmic string spacetime
\cite{Book.2000.Vilenkin}. Cosmic strings are topologically stable
gravitational defects.
According to the grand unified theories, these defects arise from a
vacuum phase transition in the near universe.
Recently, several studies have been developed in the theoretical context
\cite{GRG.2018.50.125,PRD.2018.98.063519,IJMPA.2018.33.1850158,
PRD.2017.96.024040,PRD.2018.97.085023,EPJC.2018.78.13}
and also by evidence of cosmic strings
\cite{PRL.2017.118.051301,PLB.2018.778.392,IJMPD.2018.27.1850094,
MNRAS.2018.478.1132}.
Cosmic strings are objects of studies of current interest because of the
several important applications of topological features on physics
systems in gravitation \cite{Book_Barriola_gravitation}, condensed
matter \cite{Book.2000.Vilenkin} and cosmology
\cite{Book_Shellard_1995}.

Our work is motivated by Ref. \cite{PRC.2002.65.054313}
(see also Refs. \cite{PRC.1999.59.154,PRC.1998.58.R3065}), where
the spin and  pseudospin symmetries in the relativistic mean field with
a deformed potential are investigated.
In this context, a relation between the deformed wave function and the
spherical wave function was established at the spherical limit by using
the transformation from the cylindrical coordinate into the polar
coordinate.
This relationship enables us to investigate the inclusion of cylindrical
symmetrical potentials in the Dirac equation in other scenarios, such as
the cosmic string.
One advantage of using such symmetry limits in our work is that they
allow us to decouple the first and second order differential equations
for the spinor components (each obtained in the spin symmetry and
pseudo-spin limits, respectively).

We organize the paper as follows:
In Sec. \ref{sec:sec2}, we derive the
equation that governs the dynamics of a Dirac particle with the minimal,
nonminimal and the scalar couplings in the cosmic string spacetime.
In Sec. \ref {sec:sec3}, we consider the Dirac equation written in terms
of a set of coupled differential equations.
We investigate the existence of particular solutions for the problem by
assuming that the relativistic energy of the particle is its rest energy
in both the spin and the pseudo-spin symmetries limits.
In Sec. \ref{sec:sec4}, we investigate the dynamics considering that the
energy of the particle is different from its rest energy.
To this end, we write down the Dirac equation in its quadratic form.
We obtain the energies and the corresponded wave functions and discuss
their physical validity.
In Sec. \ref{sec:sec5}, we address some particular solutions and compare
them with previous results in the literature.
Finally, the conclusions are presented in Sec. \ref{sec:sec6}.
Here, we use natural units such as  $\hbar = c =1$.

\section{The equation of motion}
\label{sec:sec2}

In this section, we derive the Dirac equation with scalar and vector
couplings to study the motion of a Dirac oscillator in the cosmic string
spacetime.
We first define the spacetime background of an idealized cosmic string
where the oscillator will move, followed by the most general
interaction, which includes the potential of the Dirac oscillator.
The interactions, however, are chosen in such a way that analytical
solutions to the Dirac equation can be obtained.

The spacetime generated by a cosmic string is described by the following
line element in cylindrical coordinates
\begin{equation}
ds^{2}=dt^{2}-dr^{2}-\alpha^{2}r^{2}d\varphi^{2}-dz^{2},  \label{metric}
\end{equation}
with $-\infty <(t,z)<\infty $, $r\geq 0$ and $0\leq\varphi\leq 2\pi$.
The parameter $\alpha $ is related to the linear mass density $\tilde{m}$ of the
string by $\alpha =1-4\tilde{m}$ and it runs in the interval $(0,1]$ and
corresponds to a deficit angle $\gamma =2\pi (1-\alpha )$.
Geometrically, the metric in Eq. (\ref{metric}) corresponds to a
Minkowski spacetime with a conical singularity \cite{SPD.1977.22.312}.

One starts by considering the free Dirac equation, i.e., in the absence of
interactions.
The interaction will be included later.
So, we have
\begin{equation}
\left( i\gamma^{\mu }\partial_{\mu }-M\right) \Psi =0,
\end{equation}
where $\Psi$ is a four-component spinorial wave function. In order to work out in the curved spacetime, we must write the Dirac gamma
matrices $\gamma^{\mu }$ in the Minkowskian spacetime (written in terms
of local coordinates) in terms of global coordinates and subsequently
include the spinor affine connection $\Gamma_{\mu }$.
In other words, we must contract $\gamma^{\mu }$ with the inverse
tetrad,
\begin{equation}
\gamma^{\mu }=e_{a}^{\mu }\gamma^{a},  \label{gmatrices}
\end{equation}
satisfying the generalized Clifford algebra
\begin{equation}
\left\{ \gamma^{\mu },\gamma^{\nu }\right\} =2g^{\mu \nu },
\end{equation}
where $(\mu ,\nu )=(0,1,2,3)$ are tensor indices and $(a,b)=(0,1,2,3)$ are tetrad indices. The matrices $\gamma^{a}=\left( \gamma^{0},\gamma^{i}\right)$ in
Eq. (\ref{gmatrices}) are the standard Dirac matrices in Minkowski
spacetime, with
\begin{equation}
\gamma^{0}=\left(
\begin{array}{rr}
\mathbbm{1} & 0 \\
0 & \mathbb{-}\mathbbm{1}
\end{array}
\right) ,~~~\gamma^{i}=\left(
\begin{array}{cc}
0 & \sigma^{i} \\
-\sigma^{i} & 0
\end{array}
\right) , \quad (i=1,2,3)  \label{standard}
\end{equation}
where $\sigma^{i}$ are the standard Pauli matrices and $\mathbbm{1}$ is the
$2\times 2$ identity matrix.
As we are interested on in a cosmic string, we need to write down the generalized Dirac equation in the curved
spacetime background with a minimal coupling.
Therefore, the relevant equation is
\begin{equation}
  \left[
    i\gamma^{\mu }(\partial_{\mu }+\Gamma_{\mu })
    -e \gamma^{\mu} A_{\mu}-M\right] \Psi =0,
\label{dirac}
\end{equation}
where $e$ is the electric charge and $A_{\mu }$ denotes the vector
potential associated with the electromagnetic field.
The spinor affine connection is often written as
\cite{APPB.2010.41.1827}
\begin{equation}
\Gamma_{\mu }=\frac{1}{8}\omega_{\mu ab}\left[ \gamma^{a},\gamma^{b}
\right] ,  \label{conn}
\end{equation}
where $\omega_{\mu ab}$ is the spin connection, given by
\begin{equation}
\omega_{\mu ab}=\eta_{ac}e_{\nu }^{c}e_{b}^{\tau }\Gamma_{\tau \mu }^{\nu
}-\eta_{ac}e_{\nu }^{c}\partial_{\mu }e_{b}^{\nu }.
 \label{spincn}
\end{equation}
In \eqref{spincn}, $\Gamma_{\tau \mu }^{\nu }$ are the Christoffel
symbols and $\eta^{ab}$ is the metric tensor.
By the means of the spin connection, we can construct a local frame
using a basis tetrad which gives the spinors in the curved spacetime.
Here, the basis tetrad $e_{a}^{\mu }$ is chosen to be
\cite{PRD.2009.79.024008}
\begin{equation}
  e_{a}^{\mu } =
  \left(
    \begin{array}{cccc}
      1 & 0 & 0 & 0 \\
      0 & \cos \varphi & \sin \varphi & 0 \\
      0 & -\sin \varphi /\alpha r & \cos \varphi /\alpha r & 0 \\
      0 & 0 & 0 & 1
    \end{array}
  \right) ,  \label{tetrad}
\end{equation}
satisfying the condition
\begin{equation}
e_{a}^{\mu }e_{b}^{\nu }\eta^{ab}=g^{\mu \nu }.
\end{equation}
Using (\ref{tetrad}), the matrices $\gamma^{\mu }$ in Eq. (\ref{dirac})
are written more explicitly as
\begin{eqnarray}
\gamma^{0} &=&e_{0}^{t}\gamma^{0}\equiv \gamma^{t},  \label{ga0} \\
\gamma^{z} &=&e_{0}^{z}\gamma^{0}\equiv \gamma^{z},  \label{gaz}
\end{eqnarray}
\begin{eqnarray}
\gamma^{1} &=&e_{a}^{1}\gamma^{a}\equiv \gamma^{r},  \label{ga1} \\
\gamma^{r} &=&e_{0}^{r}\gamma^{0}+e_{1}^{r}\gamma^{1}+e_{2}^{r}\gamma
^{2},  \notag \\
&=&\gamma^{2}\cos \varphi +\gamma^{2}\sin \varphi ,
\end{eqnarray}
\begin{eqnarray}
\gamma^{2} &=&e_{a}^{2}\gamma^{a}\equiv \frac{\gamma^{\varphi }}{\alpha r}
,  \label{ga2} \\
\gamma^{\varphi } &=&e_{0}^{\varphi }\gamma^{0}+e_{1}^{\varphi }\gamma
^{1}+e_{2}^{\varphi }\gamma^{2},  \notag \\
&=&-\gamma^{1}\sin \varphi +\gamma^{2}\cos \varphi .
\end{eqnarray}
Given the fact that the matrices in the curved space satisfy the
condition  $\nabla_{\mu}\gamma^{\mu }=0$, i.e., they are covariantly
constant, for the specific basis tetrad (\ref{tetrad}), the affine spin
connection is found to be
\begin{equation}
  \boldsymbol{\Gamma }=\left( 0,0,\Gamma_{\varphi },0\right) ,
  \label{connec}
\end{equation}
with the non-vanishing element given by
\begin{equation}
  \Gamma_{\varphi } = \frac{1}{2}\left( 1-\alpha \right)
  \gamma_{1}\gamma_{2}.
  \label{gammaphi}
\end{equation}

We are interested on including potentials with cylindrical symmetry, in
such a way the resulting system will have translational invariance
along the $z$ direction.
Then, we can discard the third direction and thus consider the Dirac
oscillator in two spacial dimensions
\cite{JPA.1989.22.817} (see also Ref. \cite{Book.1998.Strange}),
assuming $p_{z}=0$ \footnote{Otherwise, we shall have an overall phase factor of
  the kind $e^{i p_{z} z}$ in the final wave function.}.
This assumption allows us to reduce the four-component Dirac
equation (\ref{dirac}) to a two-component spinor equation.
Moreover, according to the tetrad postulated \cite{APPB.2010.41.1827},
the $\gamma^{a}$ matrices could be any set of constant Dirac
matrices.
Thus, a convenient representation is the following
\cite{PRD.1978.18.2932,NPB.1988.307.909,PRL.1989.62.1071}
\begin{equation}
  \gamma^{0}=\sigma^{z},\quad
  \beta \gamma^{1}=\sigma^{1},
  \quad\beta \gamma^{2}=s\sigma^{2},
  \label{newmatrices}
\end{equation}
where the parameter $s$, which is twice the spin value, can be introduced to
characterize the two spin states, with $s=+1$ for spin \textquotedblleft
up\textquotedblright\ and $s=-1$ for spin \textquotedblleft
down\textquotedblright .
In the representation (\ref{newmatrices}), the
matrices (\ref{ga0}), (\ref{ga1}) and (\ref{ga2}) assume the following form:
\begin{eqnarray}
\gamma^{0} &=&\beta =\sigma^{z}, \\
\beta \gamma^{r} &=&\sigma^{r}=\left(
\begin{array}{cc}
0 & e^{-is\varphi } \\
e^{is\varphi } & 0
\end{array}
\right) , \\
\beta \gamma^{\varphi } &=&s\sigma^{\varphi }=\frac{s}{\alpha r}\left(
\begin{array}{cc}
0 & -ie^{-is\varphi } \\
ie^{is\varphi } & 0
\end{array}
\right) .
\end{eqnarray}
and Eq. (\ref{gammaphi}) becomes
\begin{equation}
\Gamma_{\varphi } = -\frac{is}{2}\left( 1-\alpha \right) \sigma^{z}.
\label{newconx}
\end{equation}

Now, let us include the interactions into the Dirac equation
(\ref{dirac}).
We consider the effective potential \cite{JPA.2007.40.6427,PRC.2012.86.052201}
\begin{equation}
M\omega i\sigma^{z}\left( \beta \boldsymbol{\gamma }\cdot \mathbf{\hat{r}}
\right) r+\frac{1}{2}\left( I+\sigma^{z}\right) \Sigma (r) +
\frac{1}{2}\left( I-\sigma^{z}\right) \Delta (r) ,
\label{potential}
\end{equation}
with
\begin{eqnarray}
\Delta (r) &=&V(r) -S(r) ,  \label{delta}\\
\Sigma (r) &=&V(r) +S(r) ,  \label{sigma}
\end{eqnarray}
where
\begin{eqnarray}
V(r) &=&V_{1}(r) +V_{2}(r) =\frac{\eta
_{C_{1}}}{r}+\eta_{L_{1}}r,  \label{vv} \\
S(r) &=&S_{1}(r) +S_{2}(r) =\frac{\eta
_{C_{2}}}{r}+\eta_{L_{2}}r,  \label{vs}
\end{eqnarray}
are cylindrically symmetric scalar and vector potentials.
The first term in Eq. (\ref{potential}) represents the Dirac oscillator.
In this manner, the time-independent Dirac equation (\ref{dirac}) with
energy $E$ can be written as
\begin{equation}
H_{D}\psi =E\psi,  \label{dirac2}
\end{equation}
where $\psi $ is a two-component spinor,
\begin{align}
H_{D} = & {} \beta \boldsymbol{\gamma }\cdot \left( \mathbf{p}_{\alpha }-i
\boldsymbol{\Gamma}-iM\omega \beta \mathbf{r}\right) +\frac{1}{2}\left(
I+\beta \right) \Sigma (r)  \notag \\
& +\frac{1}{2}\left( I-\beta \right) \Delta (r) +\beta M,
\end{align}
is the Dirac Hamiltonian and
\begin{equation}
  \mathbf{p}_{\alpha }=-i\mathbf{\nabla }_{\alpha }=
  -i \left(\frac{\partial}{\partial r} \mathbf{\hat{r}}+
    \frac{1}{\alpha r} \frac{\partial}{\partial \varphi}
    \mathbf{\hat{\varphi}}\right),
\end{equation}
is the planar spatial part of the gradient operator in the metric
(\ref{metric}).

We begin the study of the particle motion by looking for first
order solutions of the Eq. (\ref{dirac2}).
For this purpose, we write the Eq. (\ref{dirac2}) as follows,
\begin{subequations}
\begin{multline}\label{P1}
  i e^{-is\varphi }
  \left[
    -\frac{\partial }{\partial r}+M\omega r
    +\frac{is}{\alpha r}\frac{\partial}{\partial\varphi}
    -\frac{\left(1-\alpha \right)}{2\alpha r}
  \right] \psi_{2} = \\
  \left[ E-M-\Sigma (r) \right] \psi_{1},
\end{multline}
and
\begin{multline}\label{P2}
  i e^{+is\varphi }
  \left[
    -\frac{\partial }{\partial r}-M\omega r
    -\frac{is}{\alpha r}\frac{\partial}{\partial\varphi}
    -\frac{\left( 1-\alpha\right) }{2\alpha r}
  \right] \psi_{1} = \\
  \left[ E+M-\Delta (r) \right] \psi_{2},
\end{multline}
\end{subequations}
and we consider the solutions as
\begin{equation}
\psi =
\begin{pmatrix}
\psi_{1} \\
\psi_{2}
\end{pmatrix}
=
\begin{pmatrix}
\sum\limits_{m}f(r) \,e^{im\varphi } \\
\sum\limits_{m}ig(r) \,e^{i\left( m+s\right) \varphi }
\end{pmatrix}
,  \label{solution}
\end{equation}
with $m=0,\pm 1,\pm 2,\pm 3,\ldots $ being the quantum angular momentum
number.
The substitution of (\ref{solution}) into (\ref{P1}) and (\ref{P2})
gives the following set of coupled differential equations:
\begin{align}
  \left( \frac{d}{dr}+s\frac{J_{\alpha }^{-}}{r}-M\omega r\right) g_{m} =
  {} &
       \left[ E-M-\Sigma (r) \right] f_{m}, \label{EG} \\
  \left( -\frac{d}{dr}+s\frac{J_{\alpha }^{+}}{r}-M\omega r\right) f_{m} =
  {} &
      \left[ E+M-\Delta (r) \right] g_{m}.
      \label{EF}
\end{align}
where
\begin{align}
  J_{\alpha }^{\pm} = {}
  & \frac{1}{\alpha }\left[ m+s\Theta^{\pm}+\frac{s}{2}(1-\alpha )\right],
\label{J+}
\end{align}
where $\Theta^{+}=0$ and $\Theta^{-}=1$.
The reason why we are using superscripts ($\pm$) in Eq.
(\ref{J+}) will be clarified in the next section.
If we consider that $\Delta \left(r\right) =0$ and $E=-M$ or $\Sigma \left(r\right) =0$ and $E=+M$, the solutions of Eqs.
(\ref{EG}) and (\ref{EF})
represent a particular solution for the problem, which is excluded from
the Sturm-Liouville problem.
In other words, such solutions would not be part of those obtained by
solving the second-order differential equation obtained from
Eq. (\ref{dirac2}). The procedure of imposing that either $\Delta \left(r\right) =0$ or $\Sigma \left(r\right) =0$ in Eqs. (\ref{EG}) and (\ref{EF}), respectively, is known in the literature as the exact limits of spin and pseudo-spin symmetries \cite{PR.2005.414.165}. These conditions are taken into account in the next section.

\section{Particular solutions and the analysis of the spin and
  the pseudo-spin symmetries}
\label{sec:sec3}

In this section, we solve the system of first-order radial
differential equations obtained in the previous section by imposing
either the exact limits of spin and pseudo-spin symmetries.
Once we find the solutions, we must verify that they are physically
acceptable solutions.
As mentioned above, the exact limit of the spin symmetry occurs when
$\Delta\left(r\right) = 0$ ($V(r) =S(r)$ in Eq. (\ref{delta})), while
that the exact limit of the pseudospin symmetry is achieved by setting
$\Sigma(r) = 0$ ($V(r) =-S(r)$ in Eq. (\ref{sigma})).
In what follows, the superscript ($+$) holds for the spin symmetry and
($-$) holds for the pseudo-spin symmetry.
In these limits, the solutions are related to the up and down components
of the spinor in Eq. (\ref{solution}), respectively.

In order to obtain the particular solutions, let us look for the bound
state solutions which obey the following normalization condition,
\begin{equation}
\int_{0}^{\infty }\left( |f_{m}(r)|^{2}+|g_{m}(r)|^{2}\right) rdr=1\;.
\label{norm}
\end{equation}
We assume $E=\pm M$, as it was mentioned above.

\subsection{The exact spin symmetry}

Here, the particular solutions for the bound states are obtained by
considering $\Delta (r) =0$ \footnote{
  After we impose the limits of symmetry, for simplicity, we use
  $\eta_{C_{1}}=\eta_{C_{2}}=\eta_{C}$ and
  $\eta_{L_{1}}=\eta_{L_{2}}=\eta_{L}$.}
along with the assumption $E=-M$ in both Eqs. (\ref{EG}) and (\ref{EF}).
Therefore, we have
\begin{align}
 \left( \frac{d}{dr}+s\frac{J_{\alpha }^{-}}{r}-M\omega r\right)
g_{m}(r) = {} & -2\left[ M+S(r) \right] f_{m}\left(
r\right) ,  \label{dfA} \\
 \left( -\frac{d}{dr}+s\frac{J_{\alpha }^{+}}{r}-M\omega r\right)
f_{m}(r) = {} &0.
 \label{dfB}
\end{align}
Their solutions are written as
\begin{align}
f_{m}(r) = {} & a_{1}r^{sJ_{\alpha }^{+}}e^{-\frac{1}{2}M\omega r^{2}},
\label{sfA} \\
g_{m}(r) = {} & r^{-sJ_{\alpha }^{-}}e^{\frac{1}{2}M\omega r^{2}}  \notag \\
& \times \left[ a_{1}\left( M\omega \right)^{-\frac{1}{2}s\left( J_{\alpha
}^{+}+J_{\alpha }^{-}\right) -\frac{3}{2}}\Gamma_{(a,b,c)}+a_{2}\right] ,
\label{sfB}
\end{align}
with
\begin{equation}
\Gamma_{(a,b,c)}=\eta_{C}\left( M\omega \right)^{\frac{3}{2}}\Gamma
_{\left( a\right) }+\eta_{L}\left( M\omega \right)^{\frac{1}{2}}\Gamma
_{\left( b\right) }+M^{2}\omega \Gamma_{\left( c\right) },  \label{GM}
\end{equation}
where
\begin{eqnarray}
\Gamma_{\left( a\right) } &=&\Gamma \left[ \frac{1}{2}s\left( J_{\alpha
}^{+}+J_{\alpha }^{-}\right) ,\,M\omega r^{2}\right] , \\
\Gamma_{\left( b\right) } &=&\Gamma \left[ \frac{1}{2}s\left( J_{\alpha
}^{+}+J_{\alpha }^{-}\right) +1,\,M\omega r^{2}\right] , \\
\Gamma_{\left( c\right) } &=&\Gamma \left[ \frac{1}{2}s\left( J_{\alpha
}^{+}+J_{\alpha }^{-}\right) +\frac{1}{2},\,M\omega r^{2}\right] ,
\end{eqnarray}
are upper incomplete Gamma functions \cite{Book.1972.Abramowitz},
$a_{1}$ and $a_{2}$ are constants.
Let us discuss the solutions (\ref{sfA}) and (\ref{sfB}).
Since $e^{-\frac{1}{2}M\omega r^{2}}$ dominates over $
r^{sJ_{\alpha }^{+}}$ for any value of $sJ_{\alpha }^{+}$, the solution
$f_{m}(r)$ in Eq. (\ref{sfA}) converges as $r\rightarrow 0$ and
$r\rightarrow \infty $.
On the other hand, as the incomplete Gamma functions $\Gamma_{(a,b,c)}$
always diverge, so $g_{m}(r)$ in (\ref{sfB}) will only converge as
$r\rightarrow 0$ if $a_{1}=0$, yielding $f_{m}(r)=0$.
The resulting solution are
\begin{equation}
\left[
\begin{array}{c}
f_{m}(r) \\
g_{m}(r)
\end{array}
\right] =a_{2}\left(
\begin{array}{c}
0 \\
1
\end{array}
\right) r^{-sJ_{\alpha }^{-}}e^{\frac{1}{2}M\omega r^{2}},\quad
\begin{cases}
 s=\pm 1, \\
 a_{1}=0.
\end{cases}
 \label{solnq}
\end{equation}
As $M\omega >0$ in (\ref{solnq}), there are no values of $sJ_{\alpha }^{-}$
for which the functions are square-integrable.
In this case, we can
therefore conclude right away that for $E=-M$ and exact spin symmetry there
is no bound state solution.

\subsection{Exact pseudo-spin symmetry}

In this case, we impose $\Sigma (r) =0$ and $E=M$ in both Eqs.
(\ref{EG}) and (\ref{EF}).
Thus, we obtain
\begin{align}
\left( \frac{d}{dr}+s\frac{J_{\alpha }^{-}}{r}-M\omega r\right) g_{m}\left(
r\right)  = {} &0, \\
\left( -\frac{d}{dr}+s\frac{J_{\alpha }^{+}}{r}-M\omega r\right) f_{m}\left(
r\right) = {} &2\left[ M+S(r) \right] g_{m}(r) .
\end{align}
Their solutions are given by
\begin{align}
f_{m}(r) = {} & b_{1}r^{sJ_{\alpha }^{+}}e^{-\frac{1}{2}M\omega r^{2}}  \notag \\
& \times \left[ b_{1}-b_{2}(-M\omega )^{\frac{1}{2}s\left( J_{\alpha
}^{-}+J_{\alpha }^{+}\right) -\frac{3}{2}}\Gamma \left( d,e,f\right) \right]
,  \label{soli1} \\
g_{m}(r)= {} & b_{2}r^{-sJ_{\alpha }^{-}}e^{\frac{1}{2}M\omega r^{2}},
\label{soli2}
\end{align}
where $b_{1}$ and $b_{2}$ are constants, and
\begin{equation}
\Gamma_{(d,e,f)}=M^{2}\omega \Gamma_{\left( d\right) }-\text{$\eta_{C}$}
\left( -M\omega \right)^{\frac{3}{2}}\Gamma_{\left( e\right) }-\text{$\eta
_{L}$}\left( -M\omega \right)^{\frac{1}{2}}\Gamma_{\left( f\right) },
\end{equation}
with
\begin{align}
\Gamma_{\left( d\right) } = {} &\Gamma \left[ \frac{1}{2}-\frac{1}{2}s\left(
J_{\alpha }^{-}+J_{\alpha }^{+}\right) ,-M\omega r^{2}\right] , \\
\Gamma_{\left( e\right) } = {} &\Gamma \left[ -\frac{1}{2}s\left( J_{\alpha
}^{-}+J_{\alpha }^{+}\right) ,-M\omega r^{2}\right] , \\
\Gamma_{\left( f\right) } = &\Gamma \left[ 1-\frac{1}{2}s\left( J_{\alpha
}^{-}+J_{\alpha }^{+}\right) ,-M\omega r^{2}\right] .
\end{align}
Again, the incomplete Gamma functions $\Gamma
_{(d,e,f)}$ in Eq. (\ref{soli1}) always diverge, so that a normalized
solution requires that $b_{2}=0$. In such a case, the function $f_{m}(r)$ is
square-integrable only for $sJ_{\alpha }^{+}\geq 0$.
The physically acceptable solution is
\begin{equation}
\left[
\begin{array}{c}
f_{m}(r) \\
g_{m}(r)
\end{array}
\right] =b_{1}r^{sJ_{\alpha }^{+}}e^{-\frac{1}{2}M\omega r^{2}}\left(
\begin{array}{c}
0 \\
1
\end{array}
\right) , \quad
\begin{cases}
sJ_{\alpha }^{+}\geq 0, \\
a_{1}=0.
\end{cases}
 \label{solacpt}
\end{equation}
Therefore, we can conclude that for the case $E=M$ along with the exact pseudo-spin
symmetry there is a bound state solution.
Here, the existence of a particular bound state solution is guaranteed
only for $M\omega>0$.
However, there are other models in the literature where this quantity
can assume any value, so that bound states solutions are allowed for
both the spin and pseudospin symmetry limits \cite{AoP.362.196.2015}.

\section{The Dirac-Pauli equation and the analysis of both the spin
  and the pseudo-spin symmetries}
\label{sec:sec4}

In this section, we study the dynamics for the case
$E \neq \pm M$.
For this purpose, it is more convenient to work with the
Eq. (\ref{dirac2}) in its quadratic form.
In our analysis, we shall see that because of the shape of the potential
(\ref{potential}), the solutions for the radial equation are given in
terms of biconfluent Heun functions and the energy levels of the
oscillator will be determined only after imposing some quantum
conditions.

To obtain the quadratic form of the Dirac equation (\ref{dirac2}), we multiply it by the matrix operator
\begin{multline}
\beta \boldsymbol{\gamma }\cdot \left( \mathbf{p}_{\alpha }-i\boldsymbol{
\Gamma }-iM\omega \beta \mathbf{r}\right) +\beta M+E+\frac{1}{2}\left( \beta
-\mathbbm{1}\right) \Sigma (r) \\
-\frac{1}{2}\left( \mathbbm{1}+\beta \right) \Delta (r),
\end{multline}
leading to
\begin{multline}
  -\nabla_{\alpha }^{2}\psi -\frac{(1-\alpha )s\sigma^{z}}
  {i\alpha^{2}r^{2}}\frac{\partial }{\partial \varphi }
  +\frac{(1-\alpha )^{2}}{4\alpha^{2}r^{2}}+M^{2}\omega^{2}r^{2}\psi \\
  -2M\omega \left\{ \sigma^{z}+\frac{s}{\alpha }\left[ \frac{1}{i}
      \frac{\partial }{\partial \varphi }-\frac{s}{2}
      \left( 1-\alpha \right) \sigma^{z}\right] \right\} \psi
  \\
  -\Sigma (r) \Delta (r) \psi
  +\left( E+M\right) \Sigma (r) \psi
  +\left( E-M\right) \Delta (r) \psi\\
 +\left( M^{2}-E^{2}\right) \psi -\frac{1}{2}i\sigma^{r}
 \left\{ \frac{d}{dr }\left[ \Sigma (r) +\Delta
     (r) \right] \right\} \psi
 \\
 -\frac{1}{2}\sigma^{\varphi }\left\{ \frac{d}{dr}
   \left[ \Sigma \left(r\right)
     -\Delta (r) \right] \right\} \psi =0,
 \label{dirac3b}
\end{multline}
where
$\nabla_{\alpha }^{2}=\partial_{r}^{2}+(1/r)\partial_{r}
+(1/\alpha^{2}r^{2})\partial_{\varphi }^{2}$
is the planar spatial part of the Laplace-Beltrami operator in the
metric (\ref{metric}).
By inserting  the solutions (\ref{solution}) into Eq. (\ref{dirac3b}), we
obtain the following set of two coupled radial differential equations of
second-order:
\begin{multline}
 -\frac{d^{2}f(r) }{dr^{2}}-\frac{1}{r}\frac{df(r)
}{dr}+\frac{\left(J_{\alpha }^{+}\right)^{2}}{r^{2}}f(r)
+M^{2}\omega^{2}r^{2}f(r) \\
-2M\omega
\left(s J_{\alpha}^{+} +1\right) f(r)
 -\Sigma (r) \Delta (r) f(r)
\\+\left(
E+M\right) \Sigma (r) f(r) +\left( E-M\right) \Delta
(r) f(r)
+\left( M^{2}-E^{2}\right) f(r) \\
+\left[ \frac{d\Delta \left(
r\right) }{dr}\right] g(r) =0,  \label{diffA}
\end{multline}
and
\begin{multline}
 -\frac{d^{2}g(r) }{dr^{2}}-\frac{1}{r}\frac{dg(r)
}{dr}+\frac{\left(J_{\alpha }^{-}\right)^{2}}{r^{2}}g(r)
+M^{2}\omega^{2}r^{2}g(r) \\
-2M\omega \left(sJ_{\alpha }^{-}-1\right) g(r)
 -\Sigma (r) \Delta (r) g(r) \\
 +\left(
E+M\right) \Sigma (r) g(r) +\left( E-M\right) \Delta
(r) g(r)
+\left( M^{2}-E^{2}\right) g(r) \\
-\left[ \frac{d\Sigma (r) }{dr}\right] f(r) \,=0.
 \label{diffB}
\end{multline}
Notice that these two equations are coupled via the last terms
and the spin and pseudospin symmetry limits uncouple them.
So, here and henceforth we employ the following approach.
For the spin symmetry limit, we solve the problem by considering the
upper component of the spinor and denotes it by $f^{+}$
(i.e., $+$ labels the spin symmetry solution) and
for the pseudospin symmetry limit,  we consider the lower component and
denotes it by $g^{-}$
(i.e., $-$ labels the pseudospin symmetry solution).

\subsection{The analysis of both the spin and the pseudo-spin symmetries}

When we take into account the exact limits of spin and symmetries in
Eqs. (\ref{diffA}) and (\ref{diffB}), each component of the spinor
satisfies
\begin{multline}
  -\frac{d^{2}f^{+}(r) }{dr^{2}}-\frac{1}{r}\frac{df^{+}\left(
r\right) }{dr}+\frac{\left( J_{\alpha }^{+}\right)^{2}}{r^{2}}f^{+}\left(
r\right) +\varpi^{2}r^{2}f^{+}(r) \\
 +\frac{{a}^{+}}{r}f^{+}(r) +b^{+}rf^{+}(r) -\left(
   k^{+}\right)^{2}f^{+}(r) = 0,  \label{diffH+}
\end{multline}
and
\begin{multline}
-\frac{d^{2}g^{-}(r) }{dr^{2}}-\frac{1}{r}\frac{dg^{-}\left(
r\right) }{dr}+\frac{\left( J_{\alpha }^{-}\right)^{2}}{r^{2}}g^{-}\left(
r\right) +\varpi^{2}r^{2}g^{-}(r)\\
 +\frac{a^{-}}{r}g^{-}(r) +b^{-}rg^{-}(r) -\left(
k^{-}\right)^{2}g^{-}(r) =  0,  \label{diffH-}
\end{multline}
where
\begin{equation}
  \left( k^{\pm}\right)^{2} = E^{2}-M^{2}+
  2M\omega \left( sJ_{\alpha}^{\pm} \pm 1\right),
\end{equation}
$\varpi =M\omega $, $a^{\pm}=2(E\pm M) \eta_{C}$ and
$b^{\pm}=2(E\pm M) \eta_{L}$.
The differential equations (\ref{diffH+}) and (\ref{diffH-}) can be
placed in an convenient mode using, respectively, the following
solutions:
\begin{align}
  f^{+}(x) = {} & {x}^{\left| J_{\alpha }^{+}\right| }
  e{^{-\frac{1}{2}({x}^{2}+\xi_{L}^{+}x)}}y^{+}(x),
  \label{solution+}\\
  g^{-}(x) = {} & {x}^{\left| J_{\alpha }^{-}\right| }
  e{^{-\frac{1}{2}({x}^{2}+\xi_{L}^{-}x)}}y^{-}(x) ,
  \label{solution-}
\end{align}
where $x= \sqrt{\varpi }r$ and ${y}^{\pm}(x) $ satisfies
\begin{multline}
  x \left[ {y}^{\pm}(x) \right]^{\prime \prime }+
  \left[ \mathbb{J}^{\pm}-2{x}^{2}-
  \xi_{L}^{\pm}x\right] \left[ {y}^{\pm}(x) \right]^{\prime}\\
 +\left[ \left( \Delta^{\pm}-\mathbb{J}^{\pm}-1\right) x-\frac{1}{2}\left(
\mathbb{J}^{\pm}\xi_{L}^{\pm}+2\xi_{C}^{\pm}\right) \right] {y}^{\pm}\left(
x\right) =  0,  \label{heunb+}
\end{multline}
where
\begin{align}
\Delta^{\pm} = & {} \frac{\left( \xi_{L}^{\pm}\right)^{2}}{4}+\frac{\left(
k^{\pm}\right)^{2}}{\varpi }, \label{Delta+-} \\
\mathbb{J}^{\pm} = & {} 2\left| J_{\alpha }^{\pm}\right| +1, \label{J+-}
\end{align}
$\xi_{C}^{\pm}={a}^{\pm}/\sqrt{\varpi }$ and $\xi_{L}^{\pm}=b^{\pm}/\sqrt{\varpi^{3}}$.
Equation (\ref{heunb+}) is a homogeneous, linear, second-order,
differential equations defined in the complex plane.
The solutions of these equations are given in terms of the biconfluent
Heun functions by \cite{Book.2010.NIST,Book.Ronveaux1995}
\begin{align}
  f^{+}( x) = {}
  &
    e{^{-\frac{1}{2}({x}^{2}+\xi_{L}^{+}x)}}
    \Big[c_{1}{x}^{\left| J_{\alpha}^{+}\right| }
    \mathit{N}^{+}\left( 2\,{\left| J_{\alpha }^{+}\right| },
    \xi_{L}^{+},\Delta^{+},2\xi_{C}^{+},x\right)
    \nonumber\\
  &
 +c_{2}{x}^{-\left| J_{\alpha }^{+}\right| }
\mathit{N}^{+}\left(
-2 {\left| J_{\alpha }^{+}\right| },\xi_{L}^{+},\Delta
^{+},2\xi_{C}^{+},\,x\right)\Big],  \label{gsolution+}\\
  g^{-}(x) = {}
  & e{^{-\frac{1}{2}({x}^{2}+\xi_{L}^{-}x)}}
    \Big[c_{1}{x}^{\left| J_{\alpha}^{-}\right| }
    \mathit{N}^{-}\left( 2\,{\left| J_{\alpha }^{-}\right| },\xi
    _{L}^{-},\,\Delta^{-},2\xi_{C}^{-},x\right)
  \nonumber\\
  &
 +c_{2}{x}^{-\left| J_{\alpha }^{-}\right| }\mathit{N}^{-}\left(
-2{\left| J_{\alpha }^{-}\right| },\xi_{L}^{-},\Delta
^{-},2\xi_{C}^{-},x\right) \Big],  \label{gsolution-}
\end{align}
where
\begin{multline}
\mathit{N}^{\pm } \left( 2\,{\left| J_{\alpha }^{\pm }\right| }
,\,\xi_{L}^{\pm },\,\Delta^{\pm },2\,\xi_{C}^{\pm },\,x\right) \\
 =\sum\limits_{q=0}^{\infty }\frac{\mathcal{A}_{q}^{\pm }\left( 2\,{
\left| J_{\alpha }^{\pm }\right| },\,\xi_{L}^{\pm },\,\Delta^{\pm
},\,2\,\xi_{C}^{\pm }\right) }{\left( 1+2\,{\left| J_{\alpha }^{\pm
}\right| }\right)_{q}}\frac{x^{q}}{q!}.  \label{frobenius}
\end{multline}
The coefficients of the series are given by
\begin{align}
  \mathcal{A}_{0}^{\pm }= {}
  & 1,  \label{A0}\\
  \mathcal{A}_{1}^{\pm }= {}
  & \frac{1}{2}\left[ 2\,\xi_{C}^{\pm }+\xi_{L}^{\pm
}\left( 1+2\,{\left| J_{\alpha }^{\pm }\right| }\right) \right] ,
\label{A1}\\
  \mathcal{A}_{q+2}^{\pm }=
  &  \left\{ \left( q+1\right) \xi_{L}^{\pm }+\frac{1
}{2}\left[ 2\xi_{C}^{\pm }+\xi_{L}^{\pm }\left( 1+2{\left|
    J_{\alpha }^{\pm }\right| }\right) \right] \right\}
    \mathcal{A}_{q+1}^{\pm }\nonumber\\
  &
 -\left( q+1\right) \left( q+1+2\,{\left| J_{\alpha }^{\pm }\right|
    }\right)\nonumber \\
  &\times
    \left[ \Delta^{\pm }-2(\,{\left| J_{\alpha }^{\pm }\right| }
-1-q)\right] \mathcal{A}_{q}^{\pm },  \label{recur}
\end{align}
and
\begin{equation}
\left( 1+2\,{\left| J_{\alpha }^{\pm }\right| }\right)_{q}=\frac{
\Gamma \left( q+2\,{\left| J_{\alpha }^{\pm }\right| }+1\right) }{
\Gamma \left( 2\,{\left| J_{\alpha }^{\pm }\right| }+1\right) }.
\end{equation}
From the recursion relation (\ref{recur}), the function
$$\mathit{N}^{\pm}( 2\,{\left| J_{\alpha }^{\pm }\right| },\,\xi_{L}^{\pm
},\,\Delta^{\pm },2\,\xi_{C}^{\pm },\,x)$$
becomes a polynomial of
degree $n$, if and only if, the two following conditions are imposed \cite{Book.Ronveaux1995,JCAM.1991.37.161}:
\begin{equation}
\Delta^{\pm }-2\left( 1+\,{\left| J_{\alpha }^{\pm }\right| }
\right) =2n,\;\;n=0,1,2,\ldots, \label{cndA}
\end{equation}
\begin{equation}
\mathcal{A}_{n+1}^{\pm }=0.  \label{cndB}
\end{equation}
In this case, the $\left(n+1\right)$th
coefficient in the series expansion is a polynomial of degree $n$ in
$2\,\xi_{C}^{\pm }$.
When $2\,\xi_{C}^{\pm }$ is a root of this polynomial, the
$\left( n+1\right) $th and subsequent coefficients cancel and the series
truncates, resulting in a polynomial form of degree $n$ for the solution $\mathit{N}^{\pm }\left( 2\,{|J_{\alpha }^{\pm }|},\,\xi_{L}^{\pm },\,\Delta
^{\pm },\,2\,\xi_{C}^{\pm },\,x\right) $.
From the condition (\ref{cndA}), we extract the following expressions
involving the energy $E_{nm}^{\pm}$:
\begin{align}
  \left(E_{nm}^{\pm}\right)^{2}-M^{2} = {}
  & 2M\omega \left[ n+\left|J_{\alpha}^{\pm}\right|-sJ_{\alpha}^{\pm}+2 \Theta^{\pm}\right]  \notag \\
  & -\frac{\eta_{L}^{2}}{M^{2}\omega^{2}}\left( E_{nm}^{\pm}\pm M\right)^{2},
\label{EnergyA}
\end{align}
We notice in  Eq. (\ref{EnergyA}) the absence of
the parameter $\eta_{C}$.
This steams from the fact that these expressions do not represent the
energies of the system in its present form.
Actually, the condition (\ref {cndB}) allows us to establish a quantum
condition that links the energy and others physical quantities,
including $\eta_{C}$ \cite{PRC.2012.86.052201,AoP.2014.347.130,JMP.2015.56.092501}.
As a result, it is possible to express the energy in terms of all
the physical parameters involved in the problem, namely, $\eta_{C}$,
$\eta_{L}$, $M$, and $\omega $.
We emphasize that that, \textit{a priori}, we are free to choose which
parameter we want to fix.
Here, such a quantum condition is established through the frequency
$\omega$ of the system.
Therefore, we now label $\omega$ as $\omega_{nm}$.
Before performing the
procedure, let us consider the solution (\ref{frobenius}) up to second-order
in $x$ of the expansion, namely,
\begin{multline}
\mathit{N}^{\pm } \left( 2\,{|J_{\alpha }^{\pm }|},\,\xi_{L}^{\pm
},\,\Delta^{\pm },\,2\,\xi_{C}^{\pm },\,x\right) =\frac{\mathcal{A}
_{0}^{\pm }}{\left( 1+2\,{|J_{\alpha }^{\pm }|}\right)_{0}} \\
 +\frac{\mathcal{A}_{1}^{\pm }}{\left( 1+2\,{|J_{\alpha }^{\pm }|}\right)
_{1}}x+\frac{\mathcal{A}_{2}^{\pm }}{\left( 1+2\,{|J_{\alpha }^{\pm }|}
\right)_{2}}\frac{x^{2}}{2!}+\ldots  \label{Heun}
\end{multline}
with
\begin{align}
\mathcal{A}_{0}^{\pm } = {} & 1,  \label{A0n} \\
\mathcal{A}_{1}^{\pm } = {} & \,\frac{1}{2}\left[ 2\,\xi_{C}^{\pm }+\xi
_{L}^{\pm }\mathbb{J}^{\pm }\right] ,  \label{A1n} \\
\mathcal{A}_{2}^{\pm } = {} &\xi_{L}^{\pm }\left[ \xi_{C}^{\pm }+\frac{1}{2}
\xi_{L}^{\pm }\mathbb{J}^{\pm }\right] +\left[ \,\xi_{C}^{\pm }+\frac{1}{2}
\xi_{L}^{\pm }\mathbb{J}^{\pm }\right]^{2},  \label{A2n}
\end{align}
Thus, Eq. (\ref{Heun}) reads
\begin{multline}
\mathit{N}^{\pm }\left( 2\,{|J_{\alpha }^{\pm }|},\,\xi_{L}^{\pm
  },\,\Delta^{\pm },\,2\,\xi_{C}^{\pm },\,x\right) =
1+\left[ \frac{\xi
_{L}^{\pm }\mathbb{J}^{\pm }+2\,\xi_{C}^{\pm }}{2\left( 2\,{\left|
J_{\alpha }^{\pm }\right| }+1\right) }\right] x  \notag \\
 +\left[ \frac{\xi_{L}^{\pm }\left( \xi_{C}^{\pm }+\frac{1}{2}\xi
_{L}^{\pm }\mathbb{J}^{\pm }\right) +\,\left( \xi_{C}^{\pm }+\frac{1}{2}\xi
_{L}^{\pm }\mathbb{J}^{\pm }\right)^{2}-2n\mathbb{J}^{\pm }}{\left( 2\,{
\left| J_{\alpha }^{+}\right| }+1\right) \left( 2\,{\left|
J_{\alpha }^{+}\right| }+2\right) }\right] x^{2}  \notag \\
 +\ldots .
\end{multline}

\begin{figure}[th]
\centering
\includegraphics[scale=0.8]{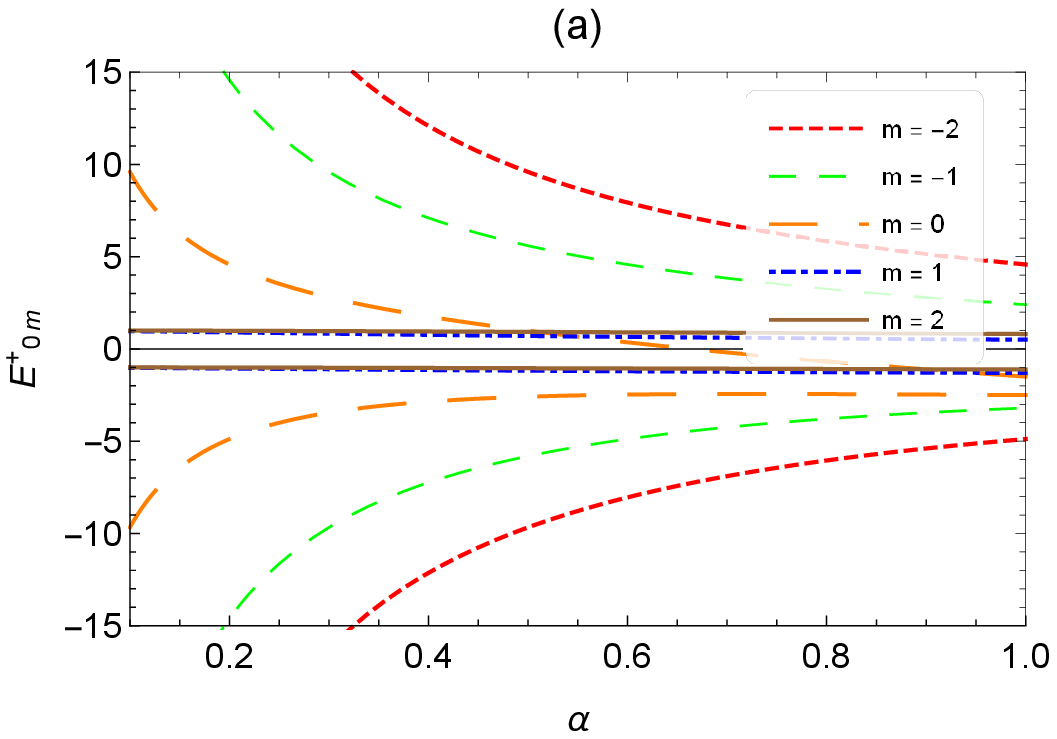}\vspace{0.3cm}
\includegraphics[scale=0.8]{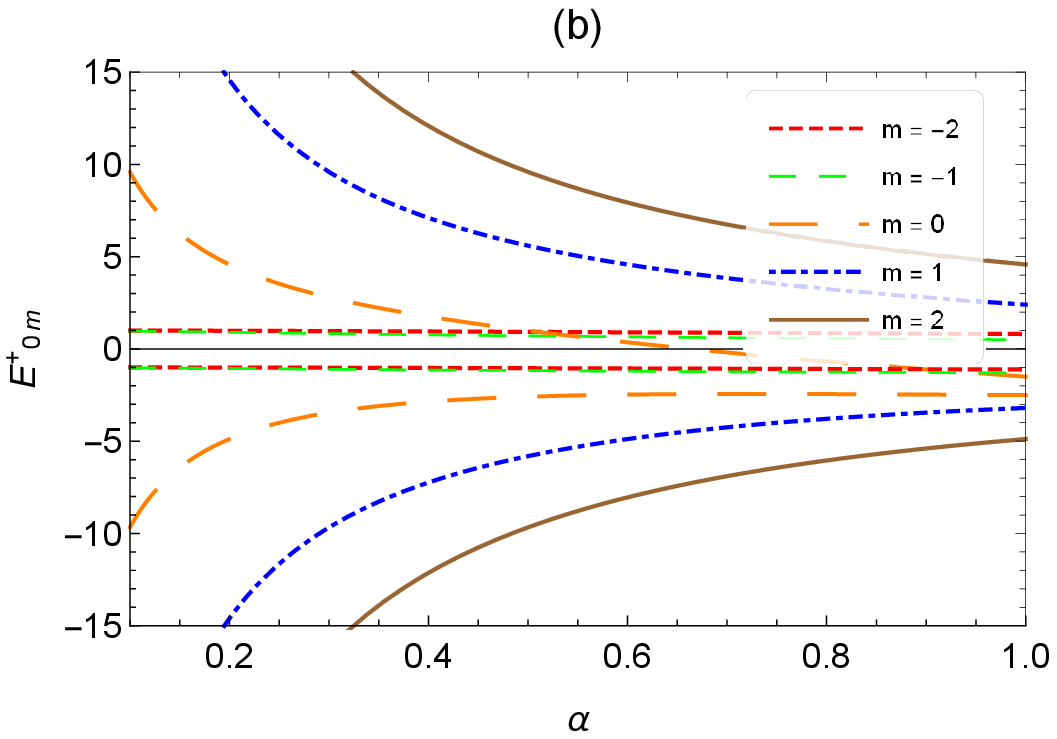}
\caption{
  (Color online) Illustration of the energy eigenvalues in the spin
  symmetry limit, $E^{+}_{0m}$, as a function of the parameter $\alpha$.
  (a) $s=1$ and (b) $s=-1$.
  We use $M=1$, $\eta_{C}=1$ and  $\protect\eta_{L}=1$.
  In (a) the energies of the states with $m<1$ become larger for $\alpha \to 0$
  whereas for $\alpha=1$ the differences between the energy levels
  decrease as well as the energy values.
  For the states with $m\geqslant 1$ (dot-dashed blue and solid brown
  lines), the energies change very slowly and are non-degenerate. In (b)
  the opposite of (a) occurs: the states with $m\geqslant 0$ are more
  energetic for $\alpha \to 0$ and less energetic for $\alpha=1$.}
\label{Plot_Energy_91a}
\end{figure}

\begin{figure}[th]
\centering
\includegraphics[scale=0.8]{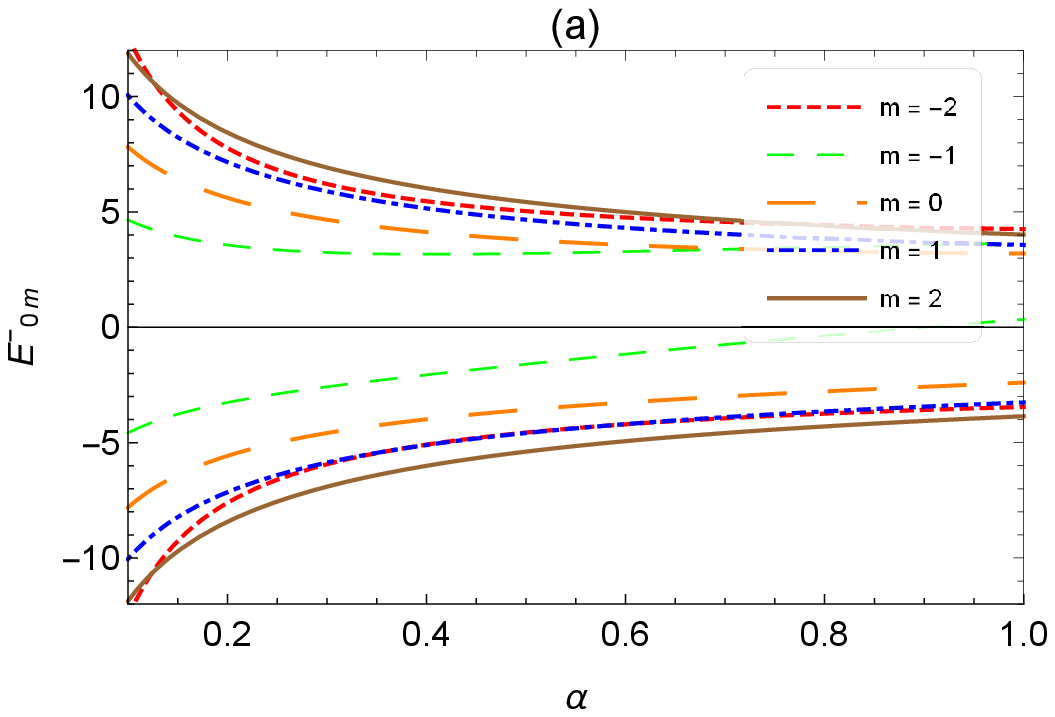}\vspace{0.3cm}
\includegraphics[scale=0.8]{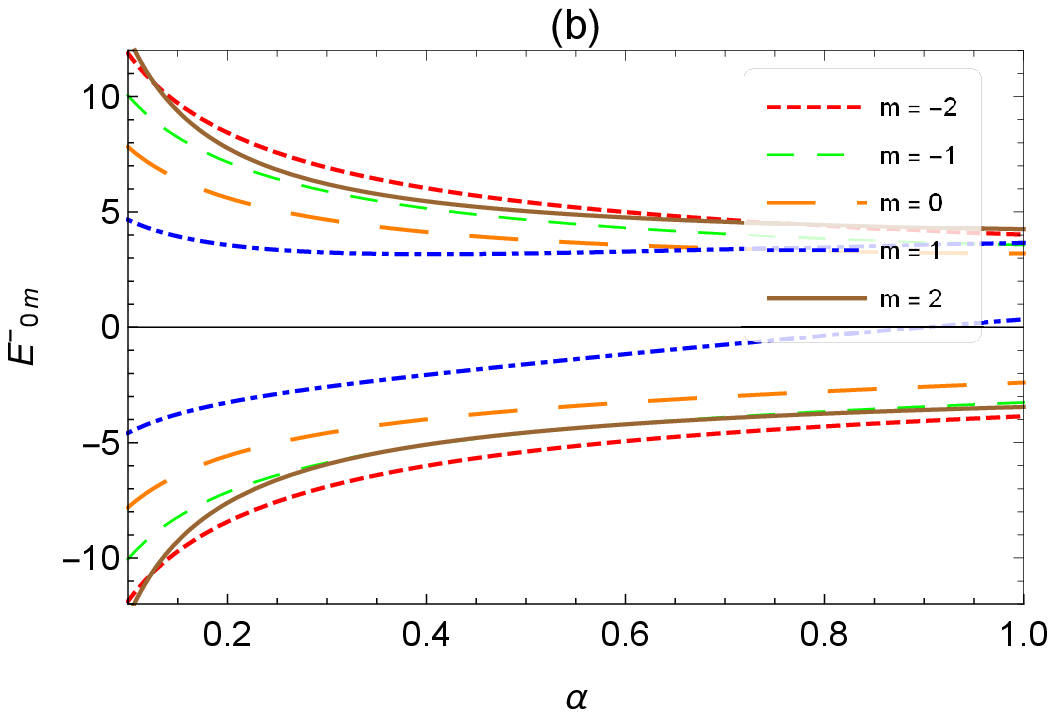}
\caption{
  (Color online) Illustration of the energy eigenvalues in the pseudospin
  symmetry limit, $E^{-}_{0m}$, as a function of the parameter $\alpha$.
  In (a) the plot for $s=1$ and (b) for $s=-1$.
  We use $M=1$, $\eta_{C}=1$ and  $\protect\eta_{L}=1$.
  The energies of the states corresponding to a given value of
  $m$ near $\alpha = 0$ in (a) and (b) are more energetic while near
  $\alpha = 1$ the differences between the energy levels decrease as
  well as their respective values.
}
\label{Plot_Energy_91b}
\end{figure}
Now let us determine the quantum condition mentioned above.
For the condition (\ref{cndB}), we must investigate
$\mathcal{A}_{n+1}^{\pm }=0$. For simplicity, we consider only the case $n=0$, which requires that
$\mathcal{A}_{1}^{\pm}=0$ in Eq. (\ref{A1n}).
This requires us to solve the equation
\begin{equation}
  2\,\frac{{a}^{\pm}}{\sqrt{\varpi }}
  +\frac{b^{\pm}}{\sqrt{\varpi ^{3}}}\mathbb{J}^{\pm}=0,
  \label{Equation_Freq_0}
\end{equation}
which provides the following frequencies related to the ground state of
the system:
\begin{equation}
  \omega _{0m}^{\pm}=
  - \frac{\eta _{L}}{2M\,\eta _{C}}\,\mathbb{J}^{\pm}.
  \label{w0mgeneral2}
\end{equation}
However, Eq. (\ref{w0mgeneral2}) will only be an acceptable quantum
condition if $\eta_{L}/\eta_{C}<0$ to ensure that the frequencies
$\omega_{0m}^{\pm}$ are positive.
Thus, respective energies corresponding to the ground state are
\begin{align}
  E_{0m}^{\pm} = {}
  & \frac{4M\,\eta_{C}^{2}}{1+\left( \mathbb{J}^{\pm}\right)^{2}}
    \left[
    \mp 1 (\pm)
    \sqrt{
    1+\frac{\left(\mathbb{J}^{\pm}\right)^{2}}{4M^{2}\eta_{C}^{2}}Q_{0m}^{\pm}}
    \right],  \label{Eq_Energy_91}
\end{align}
where
\begin{align}
Q_{0m}^{\pm} = {} &\left[ \frac{\eta_{L}}{\,\eta_{C}}\left( {\left|
J_{\alpha }^{\pm}\right| }-sJ_{\alpha }^{\pm}+2\Theta^{\pm}\right) \mathbb{J}^{\pm}-\frac{
4M^{2}\,\eta_{C}^{2}}{\left( \mathbb{J}^{\pm}\right)^{2}}+M^{2}\right]
\notag \\
&\times \left[ 1+\frac{\left( \mathbb{J}^{\pm}\right)^{2}}{4\,\eta_{C}^{2}}
\right].
\end{align}
In \eqref{Eq_Energy_91}, the notation $(\pm)$ refers to the
particle and antiparticle energies.
The energies in Eq. (\ref{Eq_Energy_91}) now
depend on all the physical parameters involved in the problem.
In Figs.  \ref{Plot_Energy_91a} and \ref{Plot_Energy_91b}, we plot the
profile of these energies as a function of the parameter $\alpha$.
In both plots we clearly see that the energy levels of the particle and
antiparticle belong to the same spectrum and, moreover, there is no
channel that allows the spontaneous creation of particles because none
of the lines of the spectrum cross each other.

\section{Particular cases}
\label{sec:sec5}

In this section, we study particular solutions of problem solved in the
previous section.
Namely, we will investigate three cases.
For the first two, the solutions of the resulting equations are given in
terms of biconfluent Heun functions whereas the third, which will not
involve scalar and vectorial interactions, will be given in terms of the
confluent hypergeometric function.

Let us then return to Eq. (\ref{heunb+}) and solve it for the particular
case $\eta_{L}=0$.
The resulting equation governs the dynamics of a two-dimensional Dirac
oscillator interacting with the potential $\eta_{C}/r$.
In this case, the solutions are given by
\begin{align}
\tilde{f}^{+}( x) = {} & \mathit{\tilde{c}}_{1}{x}
^{\left| J_{\alpha }^{+}\right| }e{^{-\frac{1}{2}{x}^{2}}}\mathit{
\tilde{N}}^{+}\left( 2{\left| J_{\alpha }^{+}\right| },0,\Delta
^{+},2\xi_{C}^{+},x\right)  \notag \\
 &+\mathit{\tilde{c}}_{2}\,{x}^{-\left| J_{\alpha
}^{+}\right| }e{^{-\frac{1}{2}{x}^{2}}}\mathit{\tilde{N}}^{+}\left(
-2\,{\left| J_{\alpha }^{+}\right| },0,\Delta^{+},2\xi
_{C}^{+},x\right),\\
\tilde{g}^{-}(x) = {} & \mathit{\tilde{c}}_{1}{x}
^{\left| J_{\alpha }^{-}\right| }e{^{-\frac{1}{2}{x}^{2}}}\mathit{
\tilde{N}}^{-}\left( 2 {\left| J_{\alpha }^{-}\right| },0,\Delta
^{-},2\xi_{C}^{-},x\right)  \notag \\
 &+\mathit{\tilde{c}}_{2}\,{x}^{-\left| J_{\alpha
}^{-}\right| }e{^{-\frac{1}{2}{x}^{2}}}\mathit{\tilde{N}}^{-}\left( -2{
\left| J_{\alpha }^{-}\right| },0,\Delta^{-},2\xi
_{C}^{-},x\right) .
\end{align}
\begin{figure}[!t]
\centering
\includegraphics[scale=0.8]{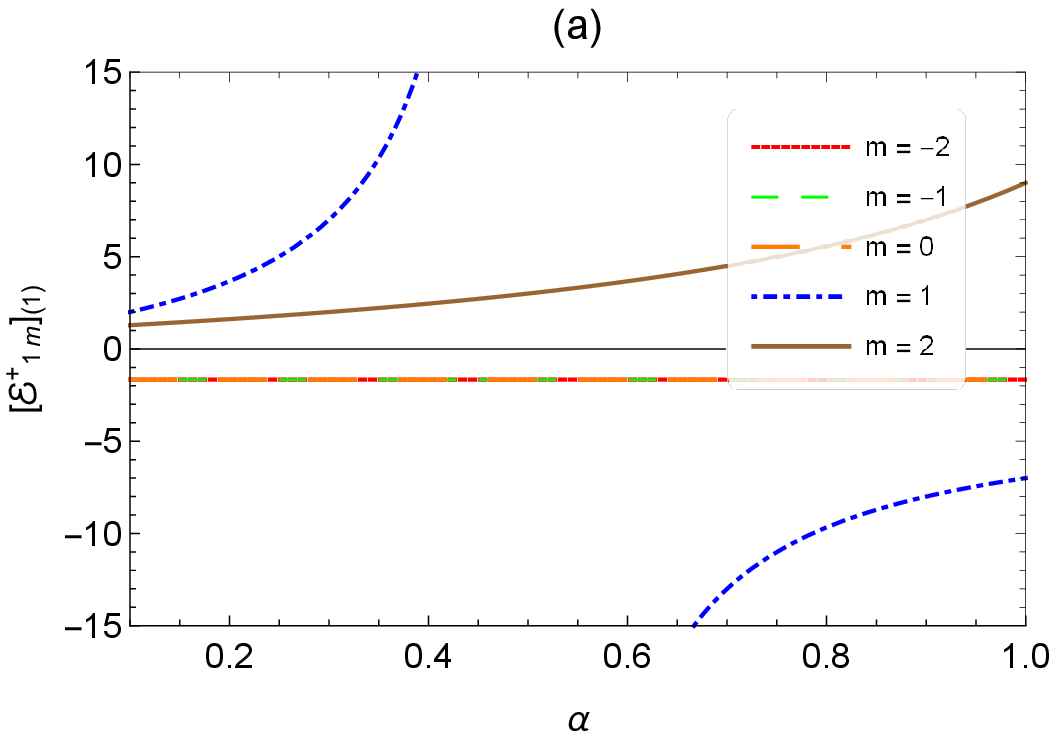}\vspace{0.3cm}
\includegraphics[scale=0.8]{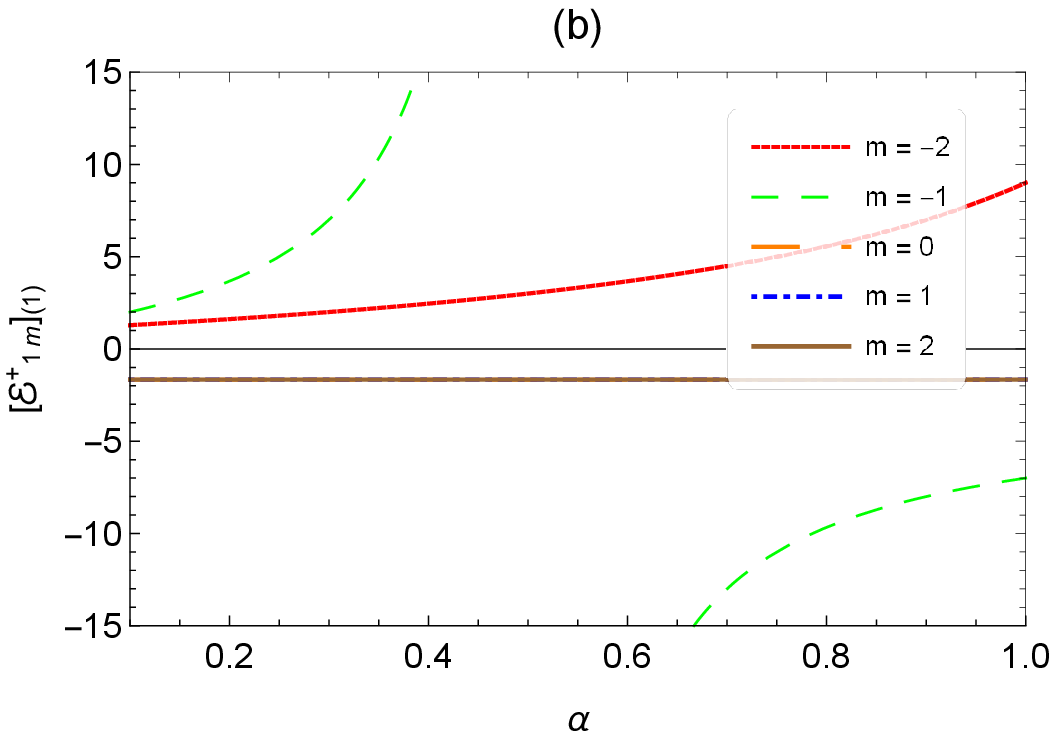}
\caption{
  (Color online)
  Illustration of the energy levels in the spin symmetry limit,
  $\left[\mathcal{E}_{1m}^{+}\right]_{p}$, as a function of the
  parameter $\alpha$ for the particular case when
  $\protect\eta_{L}=0$.
  In (a) the plot for $s=1$ and (b) for $s=-1$.
  We use $M=1$ and $\eta_{C}=1 $. In (a) the energies are degenerate for
  $m=-2,-1,0$. Energy is not defined in $\alpha = 0.5$ when $m=1$
  (dashed green line).
  The energy value for $ m = 2 $ (solid brown line) and $\alpha \to 0$
  increases while near $\alpha=1$ it decreases.
  The characteristics present in (b) are equivalent to (a) by changing
  $m$ by $-m$.}
\label{Plot_Eq_Energy_100a}
\end{figure}
\begin{figure}[!t]
\centering
\includegraphics[scale=0.8]{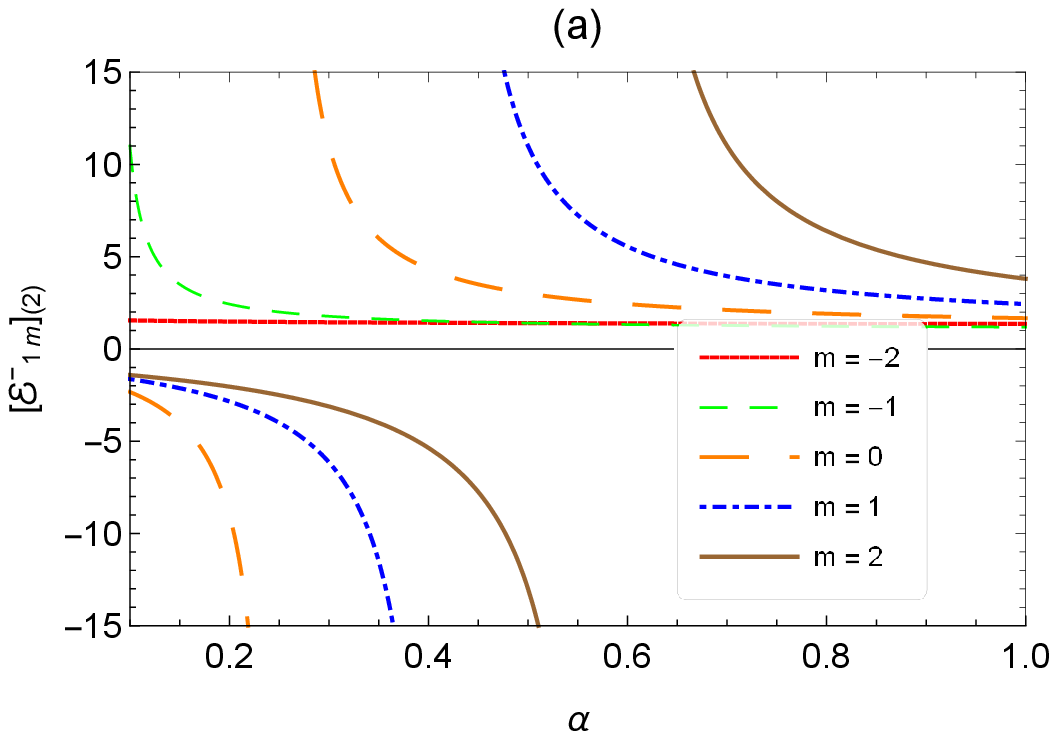}\vspace{0.3cm}
\includegraphics[scale=0.8]{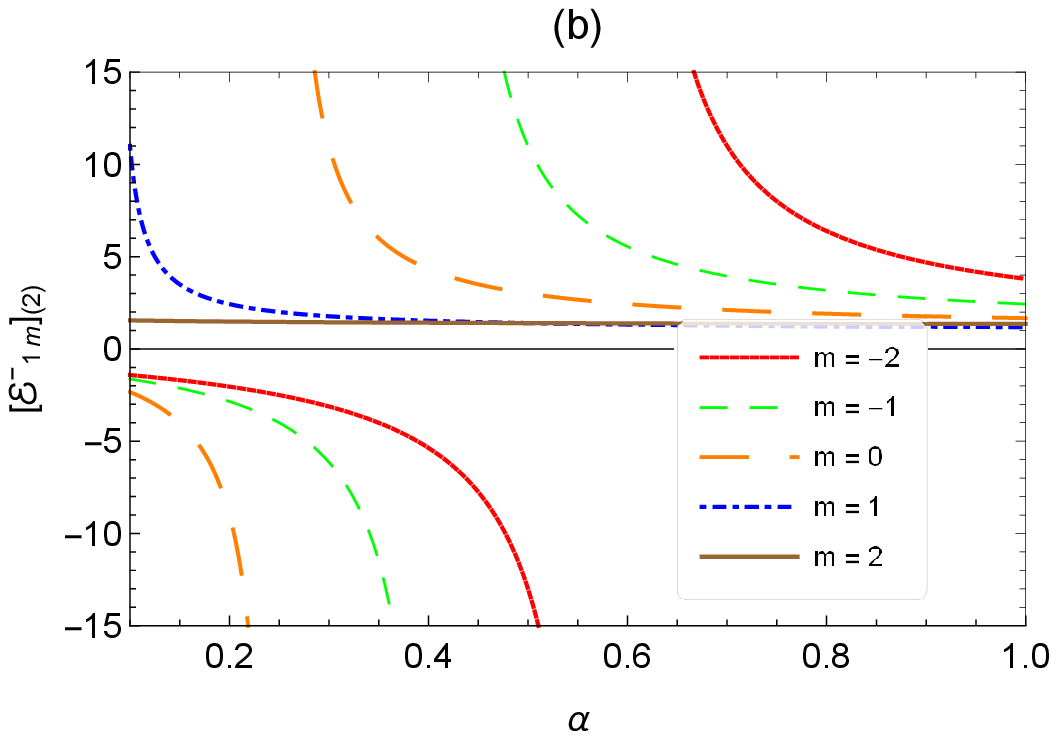}
\caption{
  (Color online) Illustration of the energy levels in the
    pseudo-spin symmetry limit, 
  $\left[\mathcal{E}_{1m}^{-}\right]_{p}$, as a function of the
  parameter $\alpha$ for the particular case when
  $\protect\eta_{L}=0$.
  In (a) the plot for $s=1$ and (b) for $s=-1$.
  We use $M=1$ and $\eta_{C}=1 $.
  In (a) the energy of the states are not defined when the
  parameter $\alpha$ is $0.25$ (dashed-long orange line), 0.42
  (dashed-dot blue line) and 0.59 (solid brown line).
  The energy of the state with $m=-2$ (dot red line) changes very slowly
  and it shows no degeneracy.
  The spectrum is more energetic near the points of singularity and less
  energetic near $\alpha=0.1$ and $\alpha=1$, respectively, except the
  $m=-1$ curve (dashed green line), which is more energetic only near
  $\alpha=0.1$.
  The characteristics manifested in (b) are equivalent to (a) by
  changing $m$ by $-m$.}
\label{Plot_Eq_Energy_101b}
\end{figure}
Then, using the condition (\ref{cndA}), we find the energies
\begin{align}
  \left(\mathcal{E}_{nm}^{\pm}\right)^{2}-M^{2}= {}
  & 2M\left( n+{\left| J_{\alpha }^{\pm}\right| }
    -sJ_{\alpha }^{\pm}+2\Theta^{\pm}\right) \tilde{\omega}_{nm}^{\pm},
    \label{Energyeta0}
\end{align}
Moreover, from condition (\ref{cndB}), we consider again
$\mathcal{A}_{n+1}^{\pm }=0$ for $n=0$, and solve it for
$\tilde{\varpi}_{0m}^{\pm }$.
One can thus verify that it is not possible to extract a physically
acceptable expression for $\tilde{\varpi}_{0m}^{\pm }$. Consequently, $n=0$ is not an allowed value for the quantum
number and we need to solve $\mathcal{A}_{n+1}^{\pm}=0$ for $n=1$.
Thus, we have
\begin{align}
  \tilde{\varpi}_{1m}^{\pm} = {}
  &\frac{2\eta_{C}^{2}}{M}
    \frac{\left(\mathcal{E}_{1m}^{\pm}+M\right)^{2}}{\mathbb{J}^{\pm}},  \label{w1b0}
\end{align}
Substituting (\ref{w1b0}) into (\ref{Energyeta0})
 and solving these equations for
$\mathcal{E}_{1m}^{\pm }$, we find
\begin{subequations}
\begin{align}
\left[ \mathcal{E}_{1m}^{+}\right]_{p} = {} &\left( \frac{1+\frac{2\eta
_{C}^{2}}{\mathbb{J}_{+}}\left( 1+\mathbb{J}_{+}-2sJ_{\alpha }^{+}\right) }{
1-\frac{2\eta_{C}^{2}}{\mathbb{J}_{+}}\left( 1+\mathbb{J}_{+}-2sJ_{\alpha
}^{+}\right) }\right) M,  \label{Eq_Energy_100a} \\
\left[ \mathcal{E}_{1m}^{+}\right]_{ap} = {} & -M,
\label{Eq_Energy_100b}
\end{align}
\end{subequations}
and
\begin{subequations}
\begin{align}
  \left[ \mathcal{E}_{1m}^{-}\right]_{p}
  & =M,
    \label{Eq_Energy_101a} \\
  \left[ \mathcal{E}_{1m}^{-}\right]_{ap}
  & =-\left( \frac{1+
\frac{2\eta_{C}^{2}}{\mathbb{J}^{-}}\left( \mathbb{J}^{-}-2sJ_{\alpha
}^{-}+5\right) }{1-\frac{2\eta_{C}^{2}}{\mathbb{J}^{-}}\left( \mathbb{J}
^{-}-2sJ_{\alpha }^{-}+5\right) }\right) M,
\label{Eq_Energy_101b}
\end{align}
where the subscripts $p$ and $ap$ refer to the energies of the particle
and antiparticle, respectively.
As we are studying the dynamics for which
$\mathcal{E}_{0m}^{\pm }\neq \pm M$, the energies
$\left[ \mathcal{E}_{1m}^{+}\right]_{ap}$ and
$\left[ \mathcal{E}_{1m}^{-}\right]_{p}$
are not allowed energies for the particle.
The profiles of the energies (\ref{Eq_Energy_100a}) and
(\ref{Eq_Energy_101b}) as a function of the parameter $\alpha$ are shown
in Figs. \ref{Plot_Eq_Energy_100a} and \ref{Plot_Eq_Energy_101b},
respectively. 
We can observe in Fig. \ref{Plot_Eq_Energy_100a}(a) ($s=+1$) the
presence of degeneracy for $m=-2,-1,0$, while in Fig.
\ref{Plot_Eq_Energy_100a}(b) ($s=-1$), the degeneracy occurs for
$m=0,1,2$.
In Fig. \ref{Plot_Eq_Energy_101b}, the spectrum of the states with
$m=-2$ (Fig. \ref{Plot_Eq_Energy_101b}(a) for $s=+1$) and with $m=2$
(Fig. \ref{Plot_Eq_Energy_101b}(b) for $s=-1$) change very slowly and
are non-degenerate.

The second particular case is when $\eta_{C}=0$.
In this case,  the system consists of a Dirac oscillator interacting
with a linear potential, $\eta_{L}r$.
Thus, the solutions of Eq. (\ref{heunb+}) is again given in terms of the
Heun functions,
\end{subequations}
\begin{align}
  \bar{f}^{+}(x) = {}
  &
    e{^{-\frac{1}{2}({x}^{2}+\xi_{L}^{+})}}
    \Big[\mathit{\bar{c}}_{1}\,{x}^{\left|
    J_{\alpha }^{+}\right| }
    \mathit{\bar{N}}^{+}\left( 2\,{\left| J_{\alpha }^{+}\right| }
                      ,\xi_{L}^{+},\,\Delta^{+},0,x\right)  \notag \\
  &
+\mathit{\bar{c}}_{2}\,{x}^{-\left| J_{\alpha
}^{+}\right| }
\mathit{\bar{N}}^{+}\left( -2\,{\left| J_{\alpha }^{+}\right| },\xi
_{L}^{+},\Delta^{+},0,x\right) \Big],
\end{align}
\begin{align}
  \bar{g}^{-}(x) = {}
  &  e{^{-\frac{1}{2}({x}^{2}+\xi_{L}^{-})}}
    \Big[\mathit{\bar{c}}_{1}\,{x}^{\left|
J_{\alpha }^{-}\right| }\mathit{\bar{N}}^{-}\left( 2\,{\left| J_{\alpha }^{-}\right| }
                      ,\xi_{L}^{-},\,\Delta_{-},0,x\right)  \notag \\
  &
+\mathit{\bar{c}}_{2}\,{x}^{-\left| J_{\alpha
}^{-}\right| }
\mathit{\bar{N}}^{-}\left( -2\,{\left| J_{\alpha }^{-}\right| },\xi
_{L}^{-},\Delta_{-},0,x\right)\Big] ,
\end{align}
and the energies are given by
\begin{align}
  \left(\bar{E}_{nm}^{\pm}\right)^{2}-M^{2}= {}
  & 2M\bar{\omega}
    \left( n+\left| J_{\alpha }^{\pm}\right|
    -sJ_{\alpha}^{\pm}+1+\Theta^{\pm}\right)\nonumber \\
& -\frac{\eta_{L}^{2}}{M^{2}\bar{\omega}^{2}}\left( \bar{E}
_{nm}^{\pm}\pm M\right)^{2},
 \label{Energyetazero}
\end{align}
Note that energies (\ref{Energyetazero}) are
identical to those given in Eq. (\ref{EnergyA}).
However, the frequency $\bar{\omega}$ is not the same.
The difference between them is just the imposition established by the
condition (\ref{cndB}).
For $n=0$, we obtain the frequencies
\begin{align}
\bar{\omega}_{0m}^{\pm} ={} &0,  \label{wzero}
\end{align}
By substituting (\ref{wzero}) into the respective
energies (\ref{Energyetazero}), we find
\begin{align}
\bar{E}_{0m}^{\pm}& =\mp M (\pm) M.
\end{align}
\begin{figure}[tbp]
\centering
\includegraphics[scale=0.8]{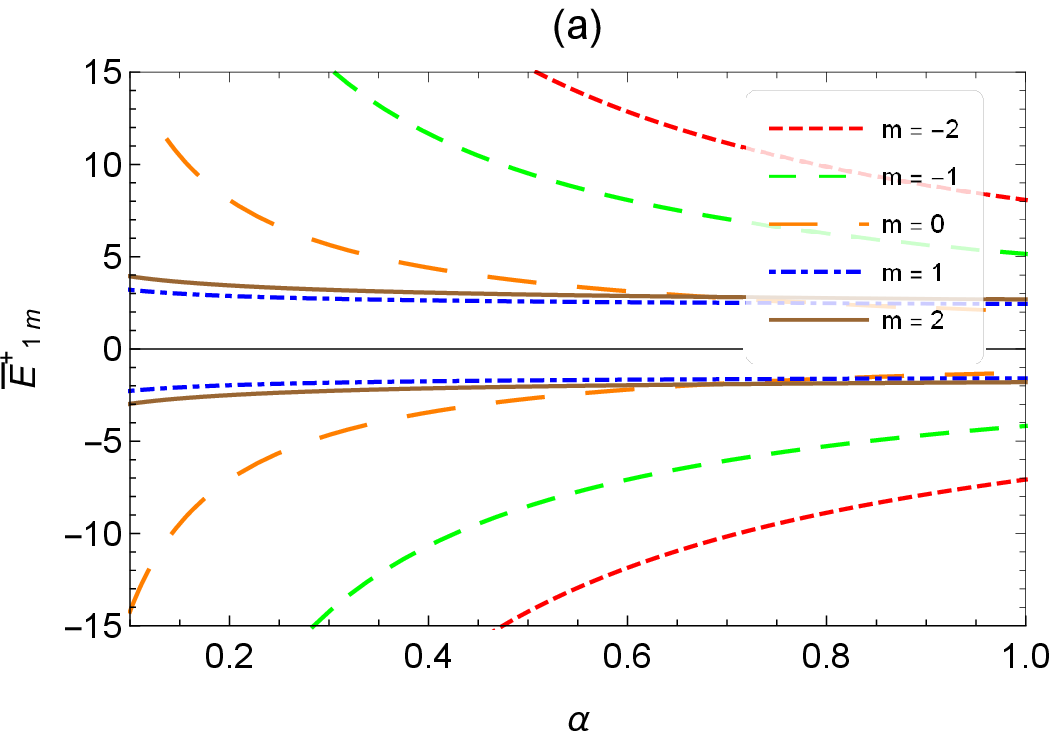}\vspace{0.3cm}
\includegraphics[scale=0.8]{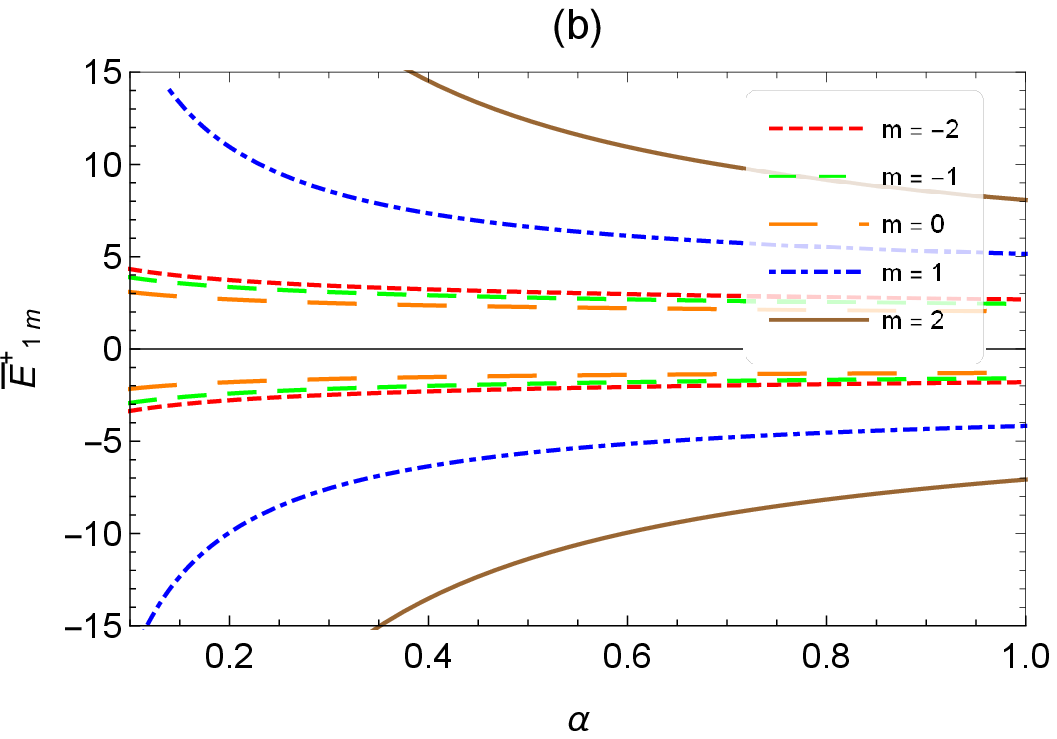}\vspace{0.3cm}
\caption{(Color online)
  The energy $\bar{E}_{1m}^{+}$ as a function of the parameter
  $\alpha$.
  (a) $s=1$, (b) $s=-1$.
  We use $M=1$ and $\eta_{L}=1$.
  In (a) the plot for $s=1$ and (b) for $s=-1$.
  We see clearly that the spectrum of the states are more energetic
  near $\alpha=0.1$ and less energetic near $\alpha=1$.
}
\label{Plot_Eq_Energy_112}
\end{figure}
\begin{figure}[tbp]
\centering
\includegraphics[scale=0.8]{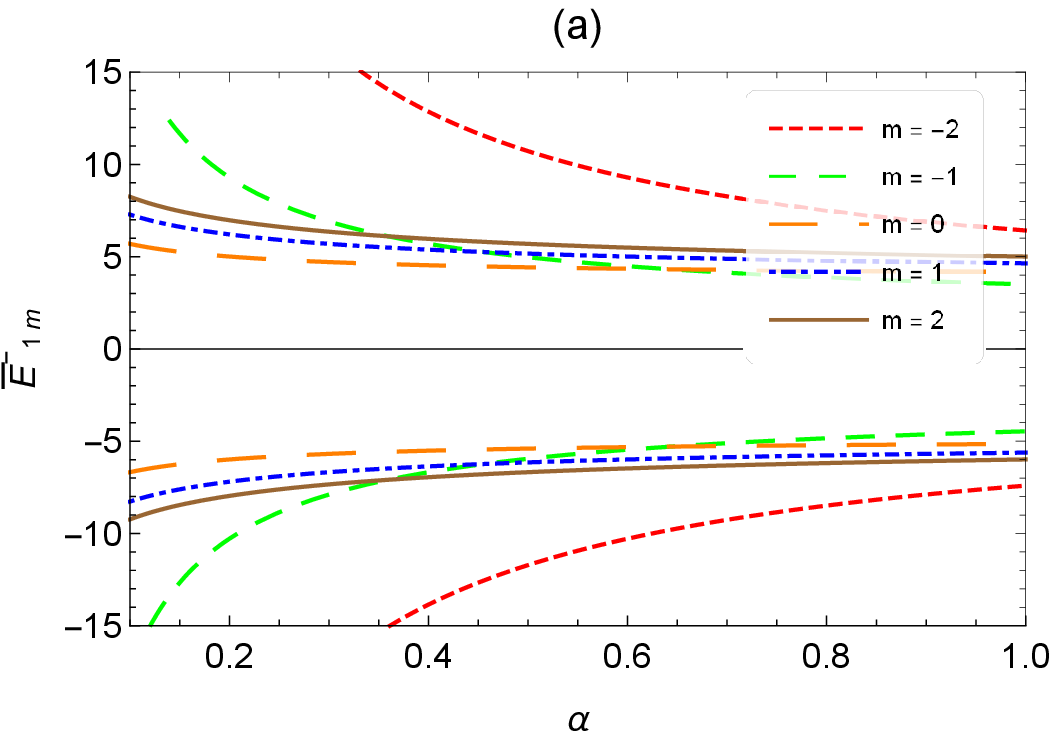}\vspace{0.3cm}
\includegraphics[scale=0.8]{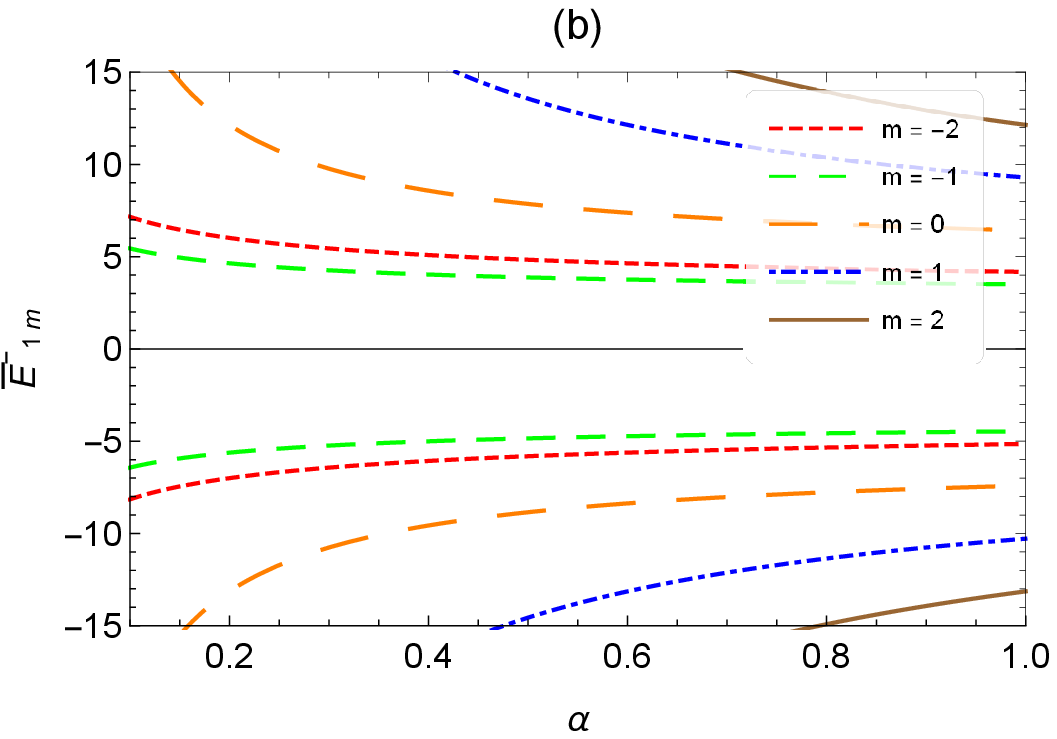}
\caption{(Color online)
  The energy $\bar{E}_{1m}^{-}$ as a function of the parameter
  $\alpha$.
  We use $M=1$ and $\eta_{L}=1$.
  In (a) the plot for $s=1$ and (b) for $s=-1$. The states are more
  energetic near $\alpha=0.1$ and less energetic near $\alpha=1$.}
\label{Plot_Eq_Energy_113}
\end{figure}
For $n=1$, we have
\begin{align}
  \bar{\omega}_{1m}^{\pm} = {}
  &\frac{1}{M}\left[ \eta_{L}^{2}\left( 1+\,\frac{\mathbb{J}^{\pm}}{2
}\right) \left( \bar{E}_{1m}^{\pm}\pm M\right)^{2}\right]^{\frac{1
}{3}},  \label{wetazero}
\end{align}
and the energies are given by
\begin{align}
\left( \bar{E}_{1m}^{\pm}\right)^{2}-M^{2} = {} & 2M\bar{\omega}_{1m}^{\pm}\left(
n+\left| J_{\alpha }^{\pm}\right| -sJ_{\alpha }^{\pm}+2\Theta^{\pm}\right)
\notag \\
& -\frac{\eta_{L}^{2}}{M^{2}\left( \bar{\omega}_{1m}^{\pm}\right)^{2}}\left(
\bar{E}_{1m}^{\pm}+M\right)^{2},  \label{Eq_Energy_112}
\end{align}
with $\bar{\omega}_{1m}^{\pm }$ given in Eq. (\ref{wetazero}).
For this particular case, it is verified that Eq. (\ref{Eq_Energy_112})
presents four energy eigenvalues being two for each type of
symmetry limit considered.
However, only two of them are physically acceptable.
The profiles of the energies $\bar{E}_{1m}^{+}$ and $\bar{E}_{1m}^{-}$
are plotted as a function of the parameter $\alpha$ for $s = 1$  and
$s = -1$ in Figs. \ref{Plot_Eq_Energy_112} and \ref{Plot_Eq_Energy_113},
respectively.
We can see that both particle and antiparticle belong to the same
spectrum and contains no degeneracy.
In Fig. \ref{Plot_Eq_Energy_112}(a), we clearly observe that the states
with $m>0$ are more affected by the curvature while in
Fig. \ref{Plot_Eq_Energy_112}(b) this occurs for the states with $m>0$.
These same characteristics are also present in
Fig. \ref{Plot_Eq_Energy_113}, the only difference is that the spacing
between each level as well as their respective energy values are
larger when compared with the spectra of the
Fig. \ref{Plot_Eq_Energy_112}.
\begin{figure}[!th!]
\centering
\includegraphics[scale=0.8]{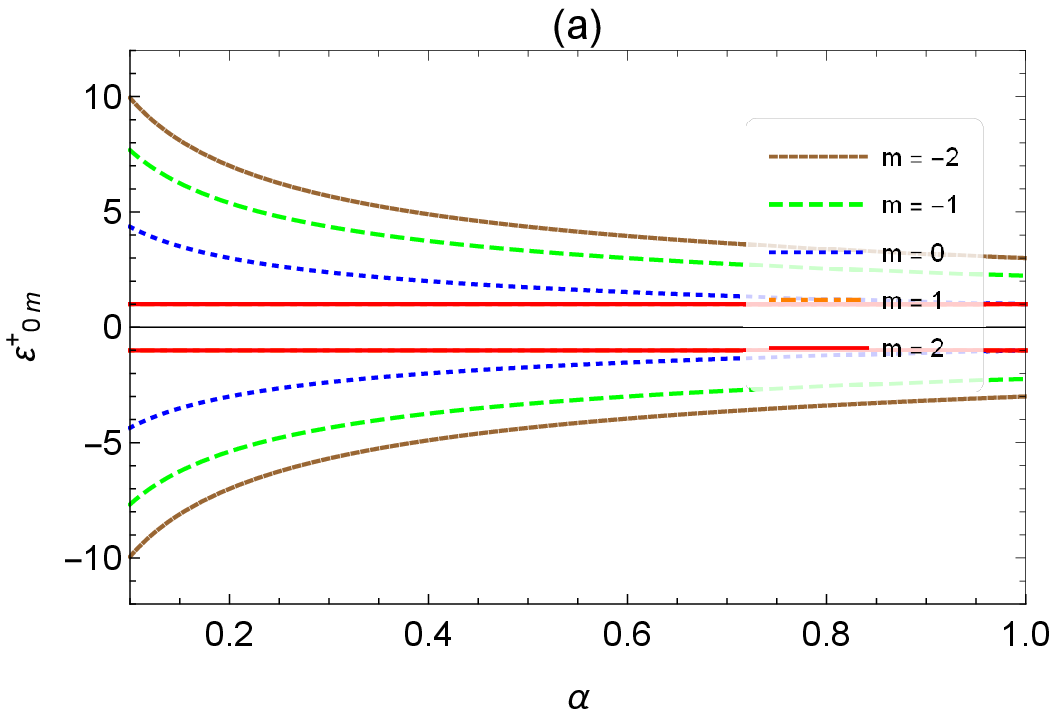}\vspace{0.3cm}
\includegraphics[scale=0.8]{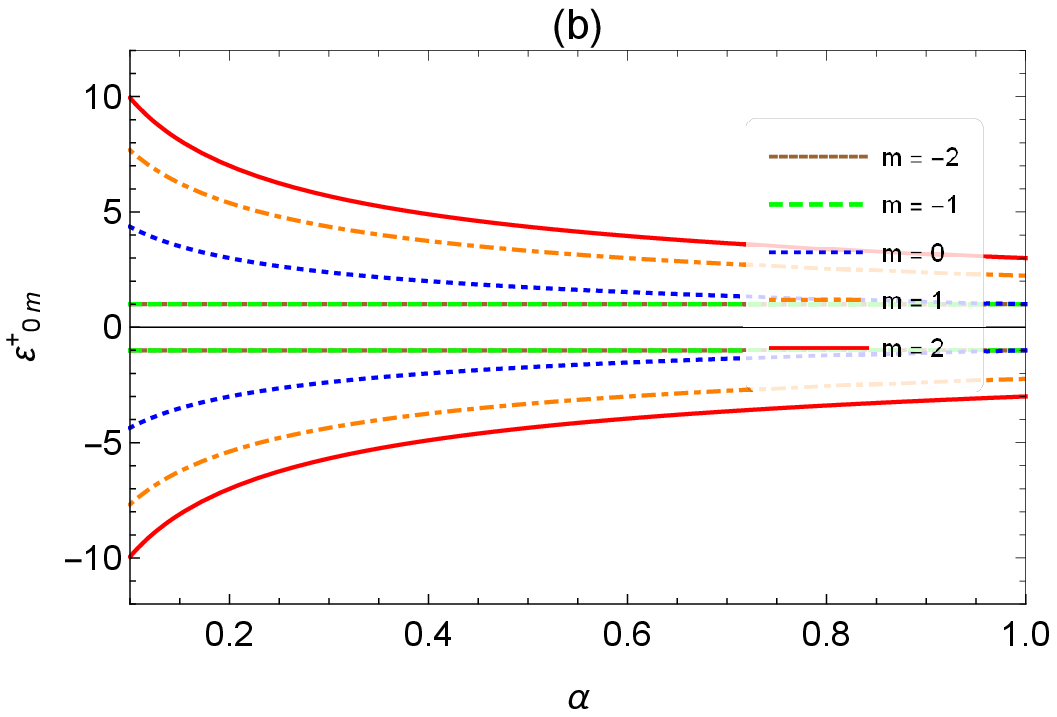}\vspace{0.3cm}
\caption{(Color online)
The energy $\epsilon_{0m}^{+}$ as a function of the parameter
  $\alpha$. In (a) the plot for $s=1$ and (b) for $s=-1$. We use $M=1$. In (a), the states with $m\leqslant0$ are more affected by curvature while the states with $m\geqslant 1$ are degenerate and are not affected by curvature. In (b), we have the situation opposite to (a): the states with $m \geqslant0$ are most affected by curvature and states with $m\leqslant -1$ are degenerate and are not affected by curvature.
}
\label{Plot_Eq_Energy_121}
\end{figure}
\begin{figure}[!th!]
\centering
\includegraphics[scale=0.8]{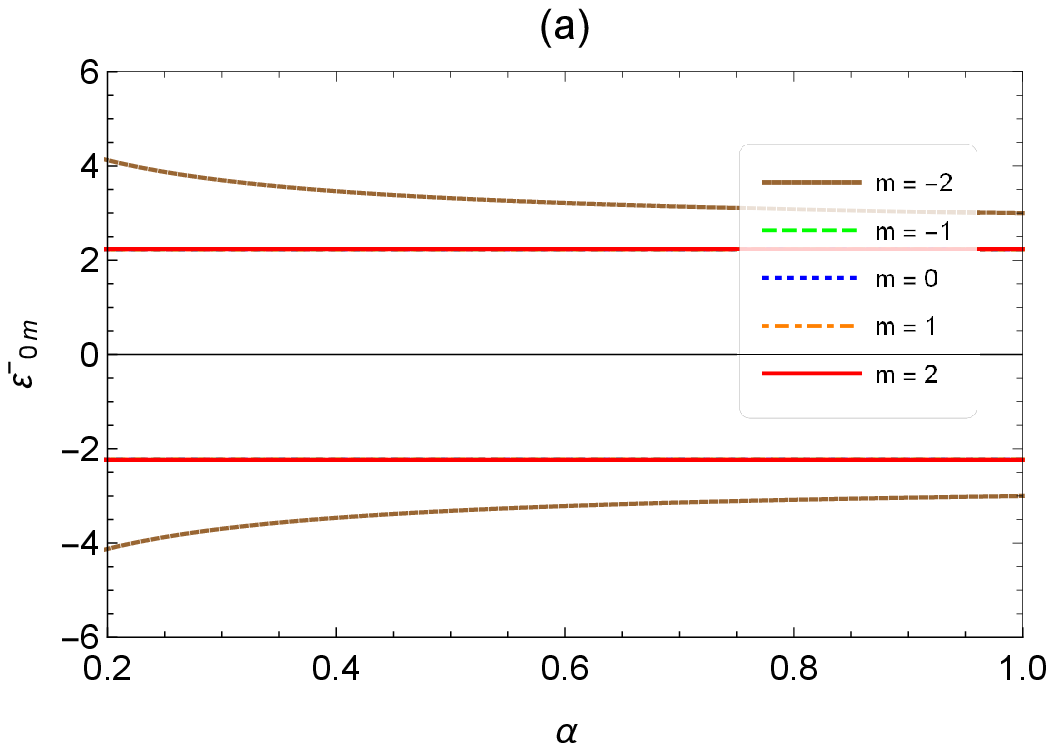}\vspace{0.3cm}
\includegraphics[scale=0.8]{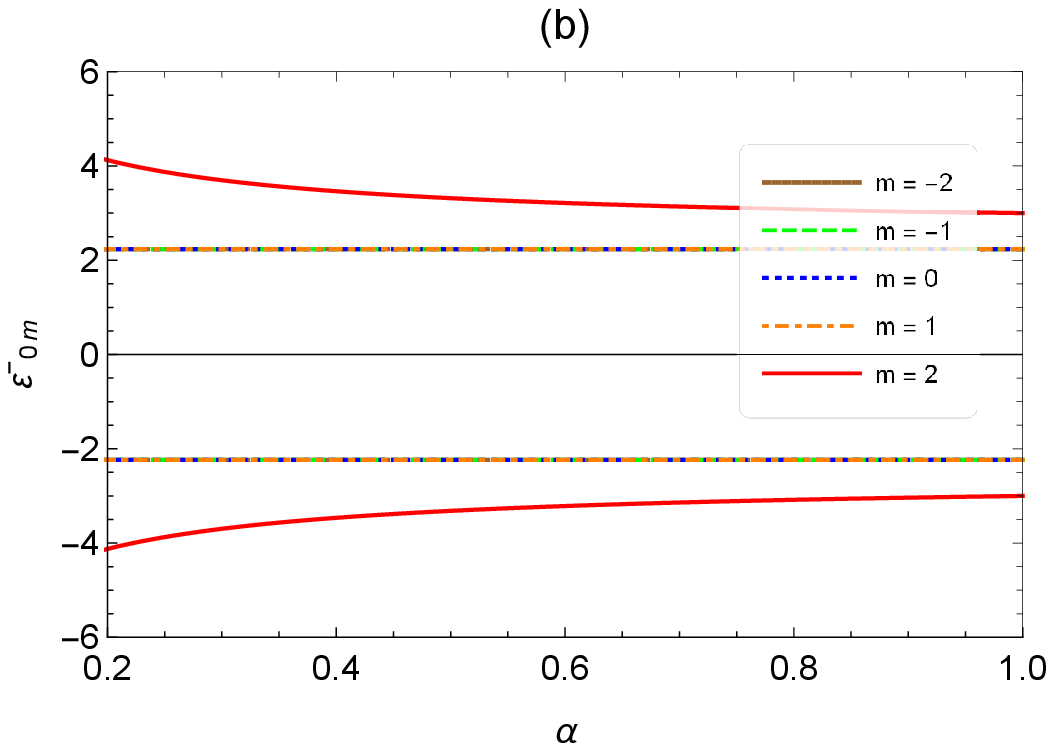}
\caption{
(Color online)
  The energy $\epsilon_{0m}^{-}$ as a function of the parameter
  $\alpha$. In (a) the plot for $s=1$ and (b) for $s=-1$. We use $M=1$.
  The only non-degenerate states affected by curvature are those with
  $m\leqslant -2$ in (a) and $m\geqslant 2$ in (b). All states with
  $m<2$ are not affected by curvature. 
}
\label{Plot_Eq_Energy_122}
\end{figure}

Finally, the last case we want to discuss in that in which
$\eta_{L}=\eta_{C}=0$ in Eq. (\ref{heunb+}).
In this case, the solutions (\ref{gsolution+}) and (\ref{gsolution-})
take the form
\begin{align}
f^{+}(x)  = {} &{x}^{\left| J_{\alpha }^{+}\right| }e{^{-
\frac{1}{2}{x}^{2}}}F^{+}(x) , \\
g^{-}(x) = {} &{x}^{\left| J_{\alpha }^{-}\right| }e{^{-
\frac{1}{2}{x}^{2}}}F^{-}(x),
\end{align}
where $x=\sqrt{\varpi }r$ and $F^{\pm }(x) $ satisfies the
Kummer differential equation \cite{Book.1972.Abramowitz,Book.2010.NIST}
\begin{multline}
\left( {F}^{\pm }\right)^{\prime \prime }(x)  +\left( \frac{
2\left| J_{\alpha }^{+}\right| +1}{x}-2{x}\right) \left( {F}^{\pm
}\right)^{\prime }(x)  \notag \\
 +\left[ \breve{\Delta}^{\pm }-\left( 2\left| J_{\alpha
}^{+}\right| +2\right) \right] \left( {F}^{\pm }\right) (x)
=0,  \label{Kummer}
\end{multline}
whose general solution is known to be
\begin{align}
  F^{\pm}(x) = {} &
  a_{n} {\mathrm{M}
    \left(
    \frac{1}{2}+
    \frac{\left| J_{\alpha }^{\pm}\right| }{2}-
    \frac{\breve{\Delta}^{\pm}}{4},
    1+\left| J_{\alpha }^{\pm}\right| ,{x}^{2}\right) }\nonumber \\
  &+b_{n} x^{-2\left| J_{\alpha }^{\pm}\right| }\mathrm{M}
    \left(\frac{1}{2}-\frac{\left|J_{\alpha }^{\pm}\right|
    }{2}-
    \frac{
    \breve{\Delta}^{\pm}}{4},1-\left| J_{\alpha }^{\pm}\right|,
    {x}^{2}\right),
\end{align}
In the above equations, $\mathrm{M}$ is the Kummer function
\cite{Book.1972.Abramowitz,Book.2010.NIST}.
For this particular case, if we write the condition (\ref{cndA}) in the
form
\begin{equation}
\frac{1}{2}+\frac{\left| J_{\alpha }^{\pm }\right| }{2}-\frac{\breve{
\Delta}^{\pm }}{4}=-n^{\prime },  \label{cndCexp}
\end{equation}
with $n^{\prime }=0,1,2,3,...$, where $\breve{\Delta}^{\pm }=\left( \breve{k}
^{\pm }\right)^{2}/M\omega $ and $\left( \breve{k}^{\pm }\right)
^{2}=\left( \epsilon_{nm}^{\pm }\right)^{2}-M^{2}+2M\omega \left(
sJ_{\alpha }^{\pm }\pm 1\right) $, the energies of the oscillator are
obtained.
Since $V(r) =S(r) =0$, spin and
pseudo-spin symmetries are now absent, and signals ($\pm $) in Eq. (\ref
{cndCexp}) are only used to represent the function $f^{+}(x) $,~$
g^{-}(x) $ (components of $\psi $ of Eq. (\ref{dirac3b}) with
positive and negative energy, respectively) of the particle.
In this way, the eigenvalues of Eq. (\ref{Kummer}) are given by
\begin{align}
\left( \epsilon_{nm}^{\pm}\right)^{2}-M^{2}& =M\omega \left[ 2n+\mathbb{J}
^{\pm}+1\right] -2M\omega \left( sJ_{\alpha }^{\pm}\pm 1\right) ,
\label{Eq_Energy_121}
\end{align}
and the unnormalized bound state wave functions are
\begin{align}
f^{+}(x) = {} &{x}^{\left| J_{\alpha }^{+}\right| }e{^{-
\frac{1}{2}{x}^{2}}\mathrm{M}\left( -n,1+\left| J_{\alpha
}^{+}\right| ,\,{x}^{2}\right) }, \\
g^{-}(x) = {} &{x}^{\left| J_{\alpha }^{-}\right| }e{^{-
\frac{1}{2}{x}^{2}}\mathrm{M}\left( -n,1+\left| J_{\alpha
}^{-}\right| ,\,{x}^{2}\right) }.
\end{align}
The energies in Eq. (\ref{Eq_Energy_121}) (for $n=0$ and $s=\pm1$) are
plotted as a function of the parameter $\alpha$ in
Figs. \ref{Plot_Eq_Energy_121} e \ref{Plot_Eq_Energy_122},
respectively.
For a particle with $s=1$ (Fig. \ref{Plot_Eq_Energy_121}(a)), all
states with $m>0$ are degenerate and are not affected by curvature while
with $s=-1$ (Fig. \ref{Plot_Eq_Energy_121}(b)), this characteristic
occurs for the states with $m<0$.
On the other hand, for an antiparticle with $s=1$
(Fig. \ref{Plot_Eq_Energy_122}(a)), only the state with $m=-2$ is
non-degenerate while with $s=-1$ (Fig. \ref{Plot_Eq_Energy_122}(b)), only
the state with $m=2$ is non-degenerate.
In Ref. \cite{EPJC.74.3187.2014}, the Dirac 2D oscillator interacting
with the Aharonov-Bohm potential in the time space of the cosmic string
was studied in the context of self-adjoint extensions.
In the absence of the Aharonov-Bohm field, the resulting equation
corresponding to the regular solution (Eq. 46) of Ref.
\cite{EPJC.74.3187.2014} reproduces the Eq. (\ref{Eq_Energy_121}).

\section{Conclusion}
\label{sec:sec6}

In this paper, we have studied the dynamics of a 2D Dirac oscillator
interacting with cylindrically symmetric scalar and vector potentials in 
the space-time of the cosmic string.
The problem was solved taking into account the spin and pseudospin
symmetry exact limits through two stages.
First we have solved the Dirac equation by looking for first order
solutions.
We used an appropriate ansatz for the Dirac equation and obtained a
system of coupled first order differential equations.
We investigated this system and verified that it admits physically
acceptable particular solutions, i.e., bound states solutions, only for
the pseudo-spin symmetry exact limit, $\Sigma =0$ and $E=M$.
In the second moment, we have constructed and solved the Dirac equation
in its quadratic form, which excludes the $E\neq \pm M$ cases from its
solutions.
For this case, we shown that the resulting radial differential equation
is the biconfluent Heun equation.
We studied the series solution of this equation as well as its
asymptotic behavior at infinity and at the origin and found two
conditions (Eqs.  (\ref{cndA}) and (\ref{cndB})) to make the series a
polynomial.
The use of these two conditions allowed us to obtain expressions for the
energies corresponding to fixed values of $n$.
In particular, we obtained the expression corresponding to the state
with $n= 0$, which is given by Eq. (\ref{Eq_Energy_91}).
We investigate how the curvature affects the energies.
For this intent, we have plotted it as a function of the parameter
$\alpha$ for each of the limits of symmetries and spin element
projection considered. 
In the case of the energy obtained for the spin symmetry limit
(Eq. (\ref{Eq_Energy_91}) with superscript $+$), we have shown that for
$s=1$ the states with $m<1$ become more energetic when $\alpha\to 0$
while for $\alpha=1$ the differences between the energy levels as well
as the respective energy values decrease.
For the states with $m\geqslant 1$, the energies change very slowly and
are non-degenerate.
When the spin element is $s=-1$, we have verified that the effects are
opposite to  those for $s=1$, namely, the states with $m\geqslant 0$ are
more energetic for $\alpha \to 0$ and  less energetic for $\alpha=1$.
These characteristics were also observed in the graph of the energies
obtained in the pseud-spin symmetry limit (Eq. (\ref{Eq_Energy_91}) with
superscript $-$).
For both $s=\pm1$, the energies of the states corresponding to a given
value of $m$ when $\alpha \to 0$ are more energetic while for
$\alpha =1$ the differences between the energy levels decrease as well
as their respective energy values. 

We also investigated some special cases for the solution of the
Eq. (\ref{dirac3b}).
In the first case, we have assumed the vanishing of the linear
potential by imposing $\eta_{L}=0$.
We obtained the energies
(Eqs. (\ref{Eq_Energy_100a})-(\ref{Eq_Energy_100b}) and
(\ref{Eq_Energy_101a})-(\ref{Eq_Energy_101b})) and plot them as a 
function of  the parameter $\alpha$ for both $s = \pm1$.
However, we have shown that the energies (\ref{Eq_Energy_100b}) and
(\ref{Eq_Energy_101a}) are not allowed. 
In the energy profile (\ref{Eq_Energy_100a}) for $s=1$, the energies of
states with $m\leqslant0$ are degenerate.
In particular, when $\alpha = 0.5$, the state energy with $m=1$ is not
defined.
We also have observed that the energy values of the state with $m=2$
when  $\alpha \to 0. $ increases while for $\alpha=1$ it decreases.
It also was verified that these same characteristics are present in the
graphic for $s=-1$.
In the plot of the energy given by Eq. (\ref{Eq_Energy_100b}) for $s=\pm1$,
other important characteristics were manifested, and these are absent in
the plot of Eq. (\ref{Eq_Energy_100a}).
For $s=1$, the energy of the states are not defined for $\alpha$ equal
to $0.25$, $0.42$ and $0.59$.
The spectrum is more energetic for $\alpha \to 0$ and $\alpha=1$, except
the $m=-1$ curve, in which is more energetic only for $\alpha \to 0$.
We have found that energy of the state with $m=-2$ changes very slowly
and are non-degenerate.
We have also found that these characteristics are present in the graphic
for $s=-1$. 

In the second particular case investigated, we have assumed $\eta_{C}=0$ and, as for the first case, four energy eigenvalues were found, but only two of them
are physically acceptable because of the requirement that $E\neq \pm
M$.
For this case, we have not found energies with a given values of $m$ and
$\alpha$ that are not allowed.
The graphs of the energies (for $s=\pm 1$) as a function of the $\alpha$
for both spin and pseudo-spin symmetry limits revealed that they are
more energetic for $\alpha \to 0$ and less energetic for  $\alpha=1.0$.
The only difference is that the spacing between the energies of the
states for a fixed $m$ in the spin symmetry limit are greater than those
in the spin symmetry limit.

In the last particular case studied, we have assumed
$\eta_{L}=\eta_{C}=0$. For this system, the resulting radial equation
was a equation Kummer differential equation type.
We obtained the energy spectrum ($\epsilon_{nm}^{\pm}$ in
Eq. (\ref{Eq_Energy_121})) and we plotted it as a function of the
$\alpha$ for both $s = \pm1$.
In the graph of the energy $\epsilon_{0m}^{+}$ for $s=1$, we have
verified that the states with $m>0$ are degenerate while for $s=-1$ this
occurs for states with $m<0$.
In the graph of the energy $\epsilon_{0m}^{-}$, we have found that only
the states with $m=-2$ (for $s=1$) and with $m=2$ (for $s=-1$) are
non-degenerate.
A feature present in all energy profiles, including the general case, is
the  absence of channel that allows creation of particles, and also no
crossings of lines, which guarantees that particle and antiparticle
belong to the same spectrum.

As a final remark, we would like to mention that the model addressed
here can be applied to other systems, especially those in condensed
matter physics.
This is due to the fact that linear defects in condensed matter, such as
disclinations and dislocations in solids, can be studied through the
same approach used to treat a cosmic string \cite{Book.2000.Vilenkin}.
A possible application would be an adaptation of the model used to
investigate how the quantum dots and antidots, with the pseudoharmonic
interaction and under the influence of external magnetic and
Aharonov-Bohm potential are influenced by the presence of a screw
dislocation as that studied in Ref. \cite{PLA.2016.380.3847} in the
context of spin and pseudo-spin symmetries.
Interesting investigations can also be made by considering non-inertial
effects on the particle dynamics \cite{PRD.1990.42.2045}.
The inclusion of non-inertial effects in relativistic and
non-relativistic quantum mechanics is an issue of current interest it
may be interesting to study some physical system in the scenario of the
problem addressed here or in some other particular geometry.

\section*{Acknowledgments}

This work was partially supported by the Brazilian agencies
Conselho Nacional de Desenvolvimento Cient\'{\i}fico e Tecnol\'{o}gico (CNPq),
Funda\c{c}\~{a}o Arauc\'aria (FAPPR),
Funda\c{c}\~{a}o de Amparo \`{a} Pesquisa de Minhas Gerais (FAPEMIG),
Funda\c{c}\~{a}o de Amparo \`{a} Pesquisa e ao Desenvolvimento
Cient\'{\i}fico e Tecnol\'{o}gico do Maranh\~{a}o (FAPEMA), and
S\~{a}o Paulo Research Foundation (FAPESP).
FMA acknowledges CNPq Grants 313274/2017-7 and 434134/2018-0, and
FAPPR Grant 09/2016.
LBC acknowledges CNPq Grants 307932/2017-6 and 422755/2018-4,
FAPEMA Grant UNVERSAL-01220/18, and FAPESP Grant 2018/20577-4.
EOS acknowledges CNPq Grants 427214/2016-5 and 303774/2016-9, and FAPEMA
Grants 01852/14 and 01202/16.
This study was financed in part by the Coordena\c{c}\~{a}o de
Aperfei\c{c}oamento de Pessoal de N\'{\i}vel Superior - Brasil (CAPES) -
Finance Code 001.

\bibliographystyle{spphys}

\end{document}